\documentclass[12pt]{iopart}
\usepackage[latin9]{inputenc}
\usepackage{amstext}
\usepackage{graphicx}
\usepackage{esint}

\makeatletter

\providecommand{\tabularnewline}{\\}

\usepackage{iopams}

\usepackage{iopams}\@ifundefined{definecolor}{\usepackage{color}}{}

\usepackage{epstopdf}

\makeatother

\begin{document}

\title{A simplified density functional theory method for charged adsorbates
on an ultrathin, insulating film supported by a metal substrate}

\author{Iván Scivetti$^{1}$ and Mats Persson$^{1,2}$}

\address{$^{1}$ Surface Science Research Centre and Department of Chemistry,
University of Liverpool, Liverpool L69 3BX, UK}

\address{$^{2}$ Department of Applied Physics, Chalmers University of Technology,
SE-412 96 Göteborg, Sweden}

\ead{scivetti@liverpool.ac.uk}

\maketitle
\date{\today} 
\begin{abstract}
A simplified density functional theory (DFT) method for charged adsorbates
on an ultrathin, insulating film supported by a metal substrate is
developed and presented. This new method is based on a previous DFT
development that uses a perfect conductor (PC) model to approximate
the electrostatic response of the metal substrate, while the film
and the adsorbate are both treated fully within DFT {[}I. Scivetti
and M. Persson, Journal of Physics: Condensed Matter \textbf{25},
355006 (2013){]}. The missing interactions between the metal substrate
and the insulating film in the PC approximation are modelled by a
simple force field (FF). The parameters of the PC model and the force
field are obtained from DFT calculations of the film and the substrate,
here shown explicitly for a NaCl bilayer supported by a Cu(100) surface.
In order to obtain some of these parameters and the polarisability
of the force field, we have to include an external, uniformly charged
plane in the DFT calculations, which has required the development
of a periodic DFT formalism to include such a charged plane in the
presence of a metal substrate. This extension and implementation should
be of more general interest and applicable to other challenging problems,
for instance, in electrochemistry. As illustrated for the gold atom
on the NaCl bilayer supported by a Cu(100) surface, our new DFT-PC-FF
method allows us to handle different charge states of adsorbates in
a controlled and accurate manner with a considerable reduction of
the computational time. In addition, it is now possible to calculate
vertical transition and reorganisation energies for charging and discharging
of adsorbates that cannot be obtained by current DFT methodologies
that include the metal substrate. We find that the computed vertical
transition energy for charging of the gold adatom is in good agreement
with experiments. \medskip{}

\end{abstract}



\section{Introduction}

A new frontier in atomic-scale science has opened up by the recent
progress in the study by scanning tunnelling microscopy (STM), non-contact
atomic force microscopy (nc-AFM) and Kelvin probe force microscopy
(KPFM) of single adsorbates on ultrathin-insulating films supported
by a metal substrate \cite{repp1,repp2,repp3,olsson,mohn,Gross09}.
Most interesting and unique properties of such films are the near
decoupling of the electronic states of the adsorbate with the electronic
states of the metal substrate \cite{repp1} and the ability to stabilise
various charge states of adsorbates on polar films, which can be switched
in a controlled manner by attachment of tunnelling electrons and holes.
These properties have been demonstrated and exploited in many experiments
including imaging of frontier orbitals \cite{repp1,repp2}, charge
state control of adsorbed species \cite{repp3,olsson}, coherent electron-nuclear
coupling in molecular wires \cite{repplilMey10} and tunnelling-induced
switching of adsorbed molecules \cite{mohn,liljeroth}.

Alongside with the on-going, exciting developments of scanning probe
microscopy experiments on these systems, density functional theory
(DFT) calculations of their electronic and geometric structure play
a crucial role in helping to unravel their physical and chemical properties
\cite{repp3,olsson,mohn}. Nevertheless, DFT calculations for these
systems are very challenging due to their system size, especially,
the large number of metal electrons, and the intrinsic self-interaction
errors in current exchange-correlation functionals \cite{DFT_lim}.
The self-interaction error and the associated delocalisation error
often result in unphysical, fractional charging of adsorbates. This
limitation not only complicates the correct identification of the
various charge states, but also leads to failures in the description
of the charge transfer process between the metal substrate and the
adsorbate. Some possible routes to surmount this limitation is offered
by DFT+U \cite{anisimov,cococcioni} or constrained DFT \cite{conDFT}.
The DFT+U approach has been applied to adatoms on insulating films
\cite{olsson} but it is not straightforward to extend this approach
to molecular adsorbates with delocalised frontier orbitals. Constrained
DFT has not so far been been applied to this problem. Regardless,
all these approaches are very challenging since they still involve
a large number of metal electrons. Replacing the metal substrate by
a positive homogeneous background was attempted in the calculation
of charged adatoms on an ultra-thin insulating film \cite{Muniz}.

In this paper we propose a new simplified and approximate DFT method
that circumvents these limitations for an adsorbate on an ultrathin
insulating film supported by a metal substrate. Here we simply assume
that the role of the metal substrate is to set the chemical potential
for the electrons, to screen the charge of adsorbates (so that the
total system is neutral), and to constrain the motion of the neighbouring
atoms in the film to the metals substrate atoms. The proposed method
is based on these assumptions and builds on our recently developed
DFT scheme \cite{ISMP}, where the insulating film and adsorbate are
treated fully within DFT and their interaction with the metal substrate
is assumed to be purely electrostatic, while the density response
of the metal surface to this interaction is treated to linear order.
Here, the metal response will simply be approximated by a classical
perfect conductor (PC) model, in which the screening charge only resides
on the image plane\cite{ISMP}. The residual interactions that are
not captured by the PC approximation will be included in a force field
(FF) between the insulating film and the metal substrate, whose parameters
are determined from DFT calculations for the insulating film and metal
substrate in absence of the adsorbate. In the development of this
force field it was required to derive appropriate corrections to the
DFT formalism (and also for the DFT-PC method) to include an external,
uniformly charged layer. Henceforth we will refer to DFT-PC when we
compute the DFT problem using the PC model and to DFT-PC-FF when using
both the PC model and the force field.\\
 With the DFT-PC-PP method the charge states of adsorbates can now
be controlled and the problem of fractional charging can be circumvented.
At this point, we would like to emphasise that the DFT-PC-FF method
will make it possible to compute transition and reorganisation energies
in charge transfer between adsorbates and the metal surface, as well
as to study the problem of excited, charge state dynamics of adsorbates.
In addition, since the electrons from the metal substrate do not appear
explicitly in the calculation, we obtain a large decrease in computational
effort, which opens up the possibility to treat large and complex
systems. 

As a specific system used to test the DFT-PC-FF method, we have considered
the case of a gold atom on a sodium chloride bilayer supported by
a copper surface. The force field was determined from DFT calculations
of a bare sodium chloride film adsorbed on the copper surface and
an external uniformly charged layer in the calculations. The adsorption
energies and the relaxed geometries of the gold atom in different
charge states that were calculated in our new DFT-PC-FF method are
compared with the results from DFT calculations of the full system.
Finally, we would like to stress that the proposed DFT-PC-FF method
is not limited to this specific system and could also straightforwardly
be extended to other interesting systems, especially those where the
ultrathin insulating film is weakly adsorbed on a metal substrate.

The paper is organised as follows. In Section \ref{sec:theory}, we
describe the theory behind the development of the DFT-PC-FF method
based on the PC model (Section \ref{sub:PC}) and how the PC model
is augmented by a force field to incorporate the missing part of the
interactions between the PC and the insulating film (Section \ref{sub:AugFF}).
This development has required an extension of periodic DFT and DFT-PC
to include an external, uniformly charged plane as described in \ref{app:ModEplan}
and \ref{app:ModEPCplan}. The computational implementation and details
are presented in Section \ref{sec:CompImplDet}. The explicit parameters
for the PC model and the force field for a sodium chloride bilayer
supported by a copper substrate are presented in Section \ref{subsec:barefilm}.
In Section \ref{subsec:chg_states}, the DFT-PC-FF method is applied
to gold adatoms in various charge states. In particular, we present
results for the transition and reorganisation energies, which cannot
be obtained from DFT calculations that explicitly include the metal
substrate. Finally, we give some concluding remarks in Section \ref{sec:conclusions}.

\section{Theory\label{sec:theory}}

We begin by introducing the perfect conductor model for charged adsorbates
on insulating film supported by a metal substrate. Here we assume
that the adsorbate is charged or discharged by the metal substrate.
The PC model only includes the mean electrostatic part of the interaction,
resulting in a net attractive interaction between the film and the
substrate. Here we develop a simple force field that captures the
residual repulsive interactions between the film and the substrate
and is augmented to the PC model. Finally, we show how the the material
specific parameters of our new scheme are obtained from DFT calculations
of the film supported by the metal substrate. Here, we will make specific
reference to a NaCl bilayer film on a Cu surface. Nevertheless, our
methodology should also be applicable to other insulating films on
various metal substrates.

\subsection{The Perfect Conductor Model for Charged Adsorbates\label{sub:PC}}

\begin{figure}
\centering %
\begin{tabular}{cc}
\includegraphics[scale=0.5]{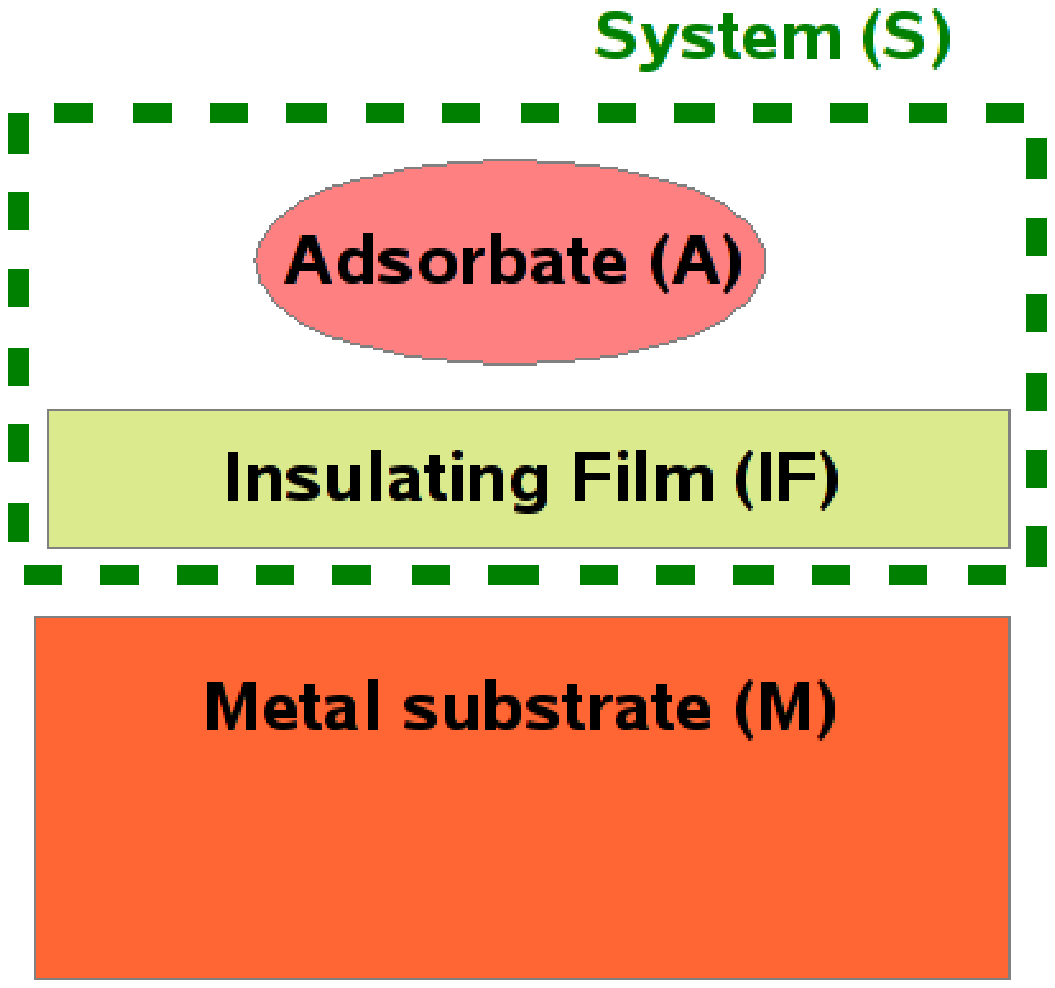}  & \hspace{0.5cm} \includegraphics[scale=0.5]{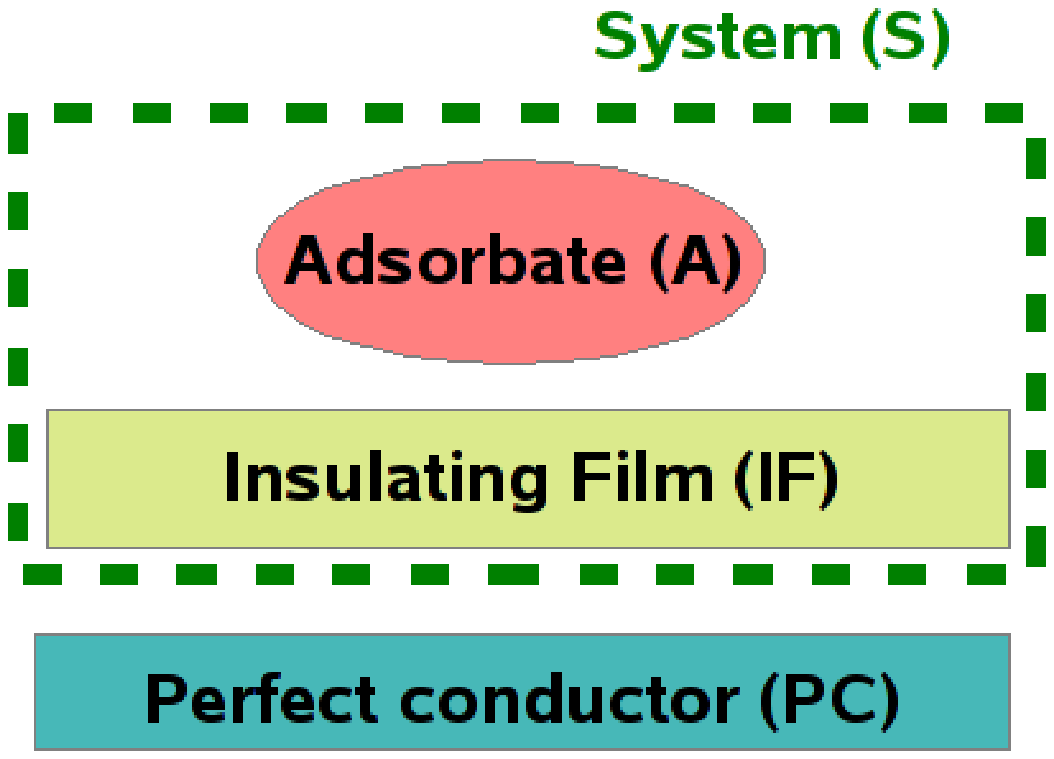}\tabularnewline
\textbf{Full System}  & \textbf{PC Approximation}\tabularnewline
\end{tabular}\caption{(colour online)(Left) Schematic representation of an adsorbate deposited
on an insulating film supported by a metal substrate. (Right) In the
perfect conductor (PC) approximation, all atoms of the metal substrate
are replaced by a simple perfect conductor model. In both figures,
we define the system S composed by the insulating film and the adsorbate,
as indicated by the dashed line.}

\label{fig:PC_replace} 
\end{figure}

As depicted schematically in Fig. \ref{fig:PC_replace} (Left), the
type of systems we will consider are composed of an adsorbate (A)
in different charge states adsorbed on an insulating film (IF) supported
by a metal substrate (M). Throughout this work, the system will be
represented in a supercell with a slab geometry of the metal substrate.
The challenge is to develop an approximation for the total energy
$E$ of the system M/IF/A based on the total energy $\bar{E}$ for
an external, charged and closed system (S) outside a metal surface.
Here S corresponds to IF/A, as schematically shown in Fig. \ref{fig:PC_replace}.
The approximate energy functional $\bar{E}$ was derived in our previous
work \cite{ISMP} using the assumption that the electron densities
$n_{s}$ and $n_{m}$ of S and M were non-overlapping, and also that
non-local contributions to the exchange-correlation functional between
S and M (such as van der Waals interactions between M and S) were
neglected. The total energy $\bar{E}$ is obtained by minimising the
following density functional,

\begin{eqnarray}
\bar{E}[n_{s}]=E_{m0}+E_{s}[n_{s}]+\int\rho_{s}({\bf r})\phi_{m0}({\bf r})d{\bf r}+\frac{1}{2}\int\rho_{ind}({\bf r})\phi_{s}({\bf r})d{\bf r}\label{eq:Ebar}
\end{eqnarray}
with respect to $n_{s}$. Here, $E_{m0}$ is the total energy of the
isolated M, $E_{s}[n_{s}]$ is the energy functional of the isolated
S, $\phi_{m0}({\bf r})$ is the unperturbed electrostatic potential
of the isolated M, $\phi_{s}({\bf r})$ is the electrostatic potential
from the charge density of S, and $\rho_{ind}({\bf r})$ is the charge
density induced by $\phi_{s}({\bf r})$ to linear order. $\bar{E}[n_{s}]$
was derived under the assumption that S was charged from or discharged
to the vacuum level but here we will now assume that the system IF/A
is charged from or discharged to the Fermi level of the metal substrate
M. Thus, we need to add an extra potential energy term to $\bar{E}[n_{s}]$,
\begin{eqnarray}
\tilde{E}[n_{s}]=\bar{E}[n_{s}]-Q_{s}\frac{\Phi}{e}\label{eq:Etilde}
\end{eqnarray}
where $Q_{s}$ is the charge of IF/A and $\Phi$ is the work function
of the isolated M/IF. Note that the charge $Q_{s}$ is now an external
parameter in Eq. (\ref{eq:Etilde}) so that different charge states
of IF/A can be treated in a controlled manner.

In the perfect conductor (PC) approximation, the explicit metal substrate
is replaced by a PC model, as depicted schematically in Fig. \ref{fig:PC_replace}
(Right). In this approximation \cite{ISMP}, $\phi_{m0}({\bf r})$
and $E_{m0}$ are set to zero and the third term of Eqn.(\ref{eq:Ebar})
vanishes. The induced charge density $\rho_{ind}({\bf r})$ in the
fourth term of Eqn.(\ref{eq:Ebar}) is localised on the PC plane and
is determined by the conditions that both the electric field and the
induced electrostatic potential inside the PC plane should be zero.
The total induced charge at the PC plane is then equal to $-Q_{s},$
so that the total charge of the supercell is zero. Note that in applying
the PC model the overlap with the electron density of IF with the
PC plane cannot be avoided and it is important to use an appropriate
expression for $\rho_{ind}({\bf r})$ on the PC plane that is valid
for overlapping densities, as discussed in Section 2.3 of Ref. \cite{ISMP}.
This density overlap and the neglected second term in Eqn.\textasciitilde{}(\ref{eq:Ebar})
make it necessary to modify the work function in Eqn.\textasciitilde{}(\ref{eq:Etilde})
so that the corresponding PC approximation of $\tilde{E}[n_{s}]$
in Eq.(\ref{eq:Etilde}) is given by, 
\begin{eqnarray}
\tilde{E}_{PC}[n_{s}]=\bar{E}_{PC}[n_{s}]-Q_{s}\frac{\Phi_{PC}}{e} & \,,\label{eq:EPCtilde}
\end{eqnarray}
 where $\bar{E}_{PC}[n_{s}]$ is the PC approximation of $\bar{E}[n_{s}]$
in Eq.(\ref{eq:Ebar}) and $\Phi_{PC}$ is the effective work function.
The procedure to determine $\Phi_{PC}$ is described in Section \ref{sub:MatSpecPar}.

\subsection{Augmentation of a Force Field to the Perfect Conductor Model\label{sub:AugFF}}

Here we develop a simple force field to approximate the energy difference,
\begin{equation}
\Delta E=E-\tilde{E}_{PC},\label{eq:DelEDef}
\end{equation}
that essentially arises from having neglected the overlapping densities
and van der Waals interactions between M and S in the PC model. We
will develop this force field for the IF using a the primitive surface
unit cell for the M/IF system. To approximate the energy difference
$\Delta E$ in Eq. (\ref{eq:DelEDef}), we have used the following
simple additive force field between the IF and the M, 
\begin{equation}
\Delta E=\sum_{k\in\mathrm{NL}}\phi_{k}(z_{\mathrm{k}})+\Delta E_{0},\label{eq:e_ForceFieldDef}
\end{equation}
where the sum of the potentials $\phi_{k}$ is over all atoms $k$
of the nearest layer (NL) of the IF to the PC plane, and $z_{k}$
is the perpendicular distance of atom $k$ from this PC plane. Here,
$\Delta E_{0}$ is a reference energy, equal to $\Delta E$ for M/IF
in its equilibrium geometry, where $z_{k}=z_{k0}$ and $\phi_{k}(z_{k0})=0$.
Note that this equilibrium geometry is determined from the total energy
$E$. This simple form of the force field is motivated in our case,
as discussed further in Section \ref{sec:results}, by (1) the interactions
between the ions in the IF are usually much stronger than their short-ranged
interactions with the M; (2) the negligible atomic relaxations of
the M even in the presence of large ionic relaxations in the IF.

In the presence of an adsorbate in different charge states, we will
use as a first approximation for $\Delta E$ in Eqn. (\ref{eq:DelEDef})
the force field of Eqn. (\ref{eq:e_ForceFieldDef}), but it will also
be corrected by making the force field polarisable, as obtained by
the introduction of a dependence of $\phi_{k}$ on the system charge
$Q_{s}$. This dependence arises from non-electrostatic interactions
of the IF with the screening charge in the M. The resulting approximate
total energy functional $E_{PC-FF}[n_{s}]$ is then given by, 
\begin{equation}
E_{PC-FF}[n_{s}]=\bar{E}_{PC}[n_{s}]-Q_{s}\frac{\Phi_{PC}}{e}+\sum_{k\in\mathrm{NL}}\phi_{k}(z_{\mathrm{k}},\sigma)+N_{\mathrm{sc}}\Delta E_{0}(\sigma)\,,\label{eq:EPC_FF}
\end{equation}
where $N_{\mathrm{sc}}$is the number of primitive surface unit cells
within the supercell and $\sigma=-eQ_{s}/N_{sc}$ is the net electron
excess of IF/A per primitive surface unit cell. Note that the forces
on the atoms in IF/A, as obtained from the Hellman-Feynman forces
generated by $\bar{E}_{PC}[n_{s}]$ and by the force field are consistent
with the energy functional $E_{PC-FF}[n_{s}]$.

\subsection{Material Specific Parameters\label{sub:MatSpecPar}}

In order to apply this approximate expression for the energy functional,
we need to determine the following material specific parameters in
the model: the perfect conductor plane position $z_{PC}$, the effective
work function $\Phi_{PC}$, the potentials $\phi_{k}(z_{\mathrm{k}},\sigma)$
in the force field and the reference energy $\Delta E_{0}(\sigma)$.
Here, we use the classical image plane position $z_{im}$ for $z_{PC}$.
According to Lang and Kohn\cite{kohn_lang1,kohn_lang2,kohn_lang3},
$z_{im}$ is determined by the linear density response of the conduction
electrons in the bare metal surface to an external homogeneous electric
field. To obtain this density response we have used a slab that represents
the metal surface in a supercell, as described and calculated explicitly
for the Cu(100) surface in \ref{app:image_plane}. Clearly, the position
of the image plane will depend on the metal substrate and its orientation. 

The difference between the effective PC workfunction $\Phi_{PC}$
and the work function $\Phi$ of the isolated M/IF is due the overlap
of the electron density of IF with the PC plane, which gives rise
to a potential difference between the PC plane and the vacuum level
and is readily obtained from the calculated electrostatic potential.

Here, the non-electrostatic interactions of the IF with the adsorbate-induced
screening charge in the M that gives rise to dependence of $\Delta E_{0}(\sigma)$
and the potentials $\phi_{k}(z_{\mathrm{k}},\sigma)$ on $\sigma$
has been estimated from calculations of $E$ and $\tilde{E}_{PC}$
for M/IF and PC/IF, respectively, where this screening charge is approximated
by the one obtained from an external, uniformly charged plane with
charge $-e\sigma$. These calculations have required us to extend
DFT and also DFT-PC to include such a charged plane in a supercell
geometry. This extension with corresponding modifications of $E$
and $\tilde{E}_{PC}$ are described in \ref{app:ModEplan} and \ref{app:ModEPCplan}.
Details of this implementation in the VASP code are presented in \ref{app:compimp}.
In these calculations, the reference geometries of M/IF and PC/IF
are determined by the equilibrium geometry obtained from the DFT calculations
for a surface primitive cell of M/IF . Similarly, the potentials $\phi_{k}(z_{\mathrm{k}},\sigma)$
in Eq.(\ref{eq:EPC_FF}) are obtained by calculating $E$ and $\tilde{E}_{PC}$
in the presence of this charged plane as a function of $z_{\mathrm{k}}$
by keeping all other atoms than atom $k$ fixed at the reference geometry.

\section{Computational implementation and details\label{sec:CompImplDet}}

All the DFT computations in this work including those based on the
PC model have been performed using the plane wave code VASP \cite{vasp}.
The implementation of the PC model in VASP has been described in our
previous work \cite{ISMP}, whereas the DFT implementation of a system
interacting with an external charged plane is described in \ref{app:compimp}.
The electron-ion interactions were handled using the projector augmented
wave method (PAW) \cite{paw} and the electronic exchange and correlation
effects were treated using the optB86b version \cite{klimes1,klimes2}
of the van der Waals density functional. The plane wave cut-off energy
was set to 400 eV.

\begin{figure}
\centering \includegraphics[scale=0.12]{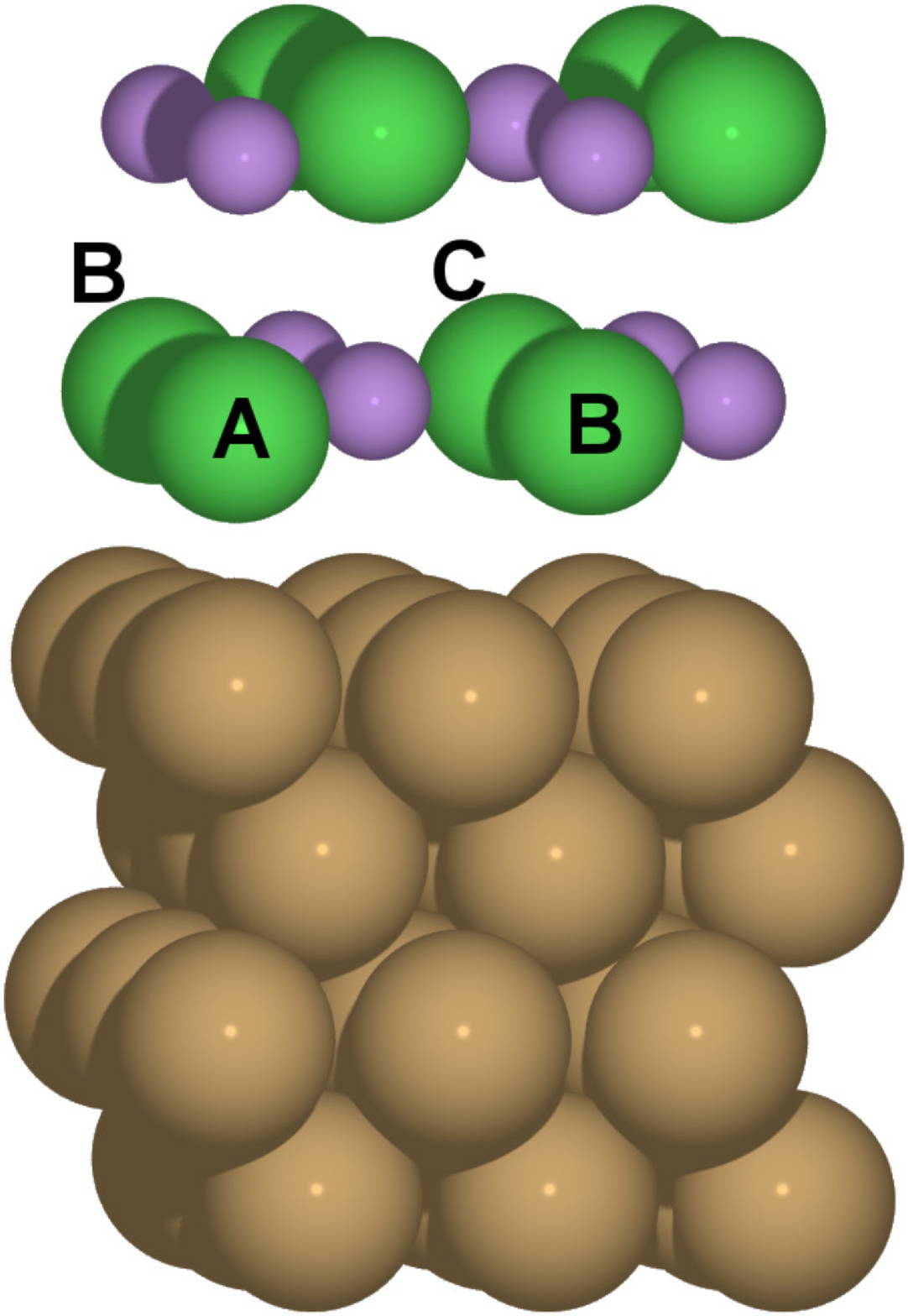}\hspace{2cm}
\includegraphics[scale=0.24]{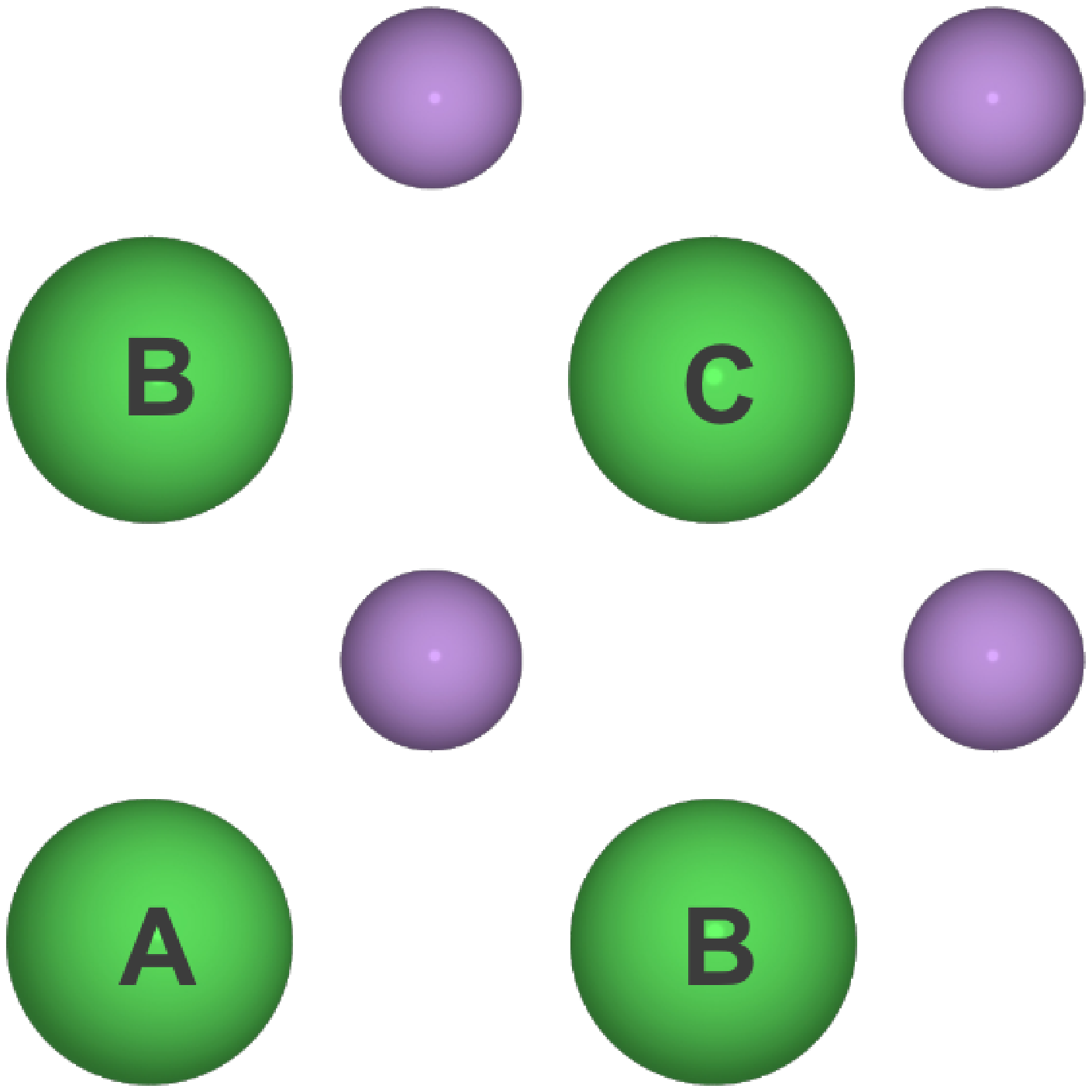} \caption{(colour online)(Left) Slab representation of a primitive surface unit
cell of a NaCl bilayer supported by a Cu(100). (Right) Top view of
the nearest NaCl layer to the Cu surface. Chlorine atoms (green) are
labelled with letters that identify their different environments with
respect to the Cu(100) substrate. In contrast, all Na atoms (violet)
were found to be equivalent with respect to the substrate atom environment.}

\label{fig:PC_cell_fit} 
\end{figure}

Given the close match 2:3 of the NaCl and Cu lattice constants, the
NaCl film is nearly commensurate with the Cu surface. Accordingly,
the corresponding primitive surface unit cell of the NaCl bilayer
supported by a Cu(100) substrate the system consists of two NaCl layers,
each layer with four Na and Cl atoms, whereas the Cu(100) substrate
is modelled by four layers with 9 Cu atoms in each layer, as shown
in Figure \ref{fig:PC_cell_fit} (Left). The interatomic distances
of the Cu atoms in the two fixed bottom layers were kept at the calculated
bulk distances of 2.546 \AA{} \cite{endnote1}. 

In the calculations behind the determination of the force field, the
supercell contained a single surface unit cell and the Brillouin zone
was sampled by $4\times4\times1$ k-points. For the calculations of
the neutral and charged Au adatom, we have used supercells containing
$2\times2$ and $3\times3$ surface unit cells, with Brillouin zone
sampling of $2\times2\times1$ and $1\times1\times1$ k-points, respectively.
All ionic relaxations were carried out until the magnitude of the
forces were smaller than 0.02 eV/\AA{}. In the calculations using
an external, uniformly charge plane, the position of this plane was
set at an average distance of 2.8 \AA{} with respect to the top NaCl
layer.

\section{Results}

\label{sec:results} We start by determining the material specific
parameters and the force field in the DFT-PC-FF method based on DFT
calculations of a bare NaCl bilayer on a Cu(100) surface. The derived
force field is then used with the DFT-PC-FF method to compute the
total energy and geometric structure of neutral and negatively charged
states of a Au adatom. In particular, we are now able to calculate
transition annd reorganistion energies.

\subsection{NaCl bilayer supported by a Cu(100) surface: force field}

\label{subsec:barefilm} The equilibrium geometry for the primitive
surface unit cell of the NaCl bilayer on the Cu(100) surfaces (Fig
\ref{fig:PC_cell_fit} (Left)) is determined from DFT calculations
including the metal substrate. From the relaxed ionic positions of
the nearest NaCl layer to the Cu(100) substrate, we find that the
four sites for the Na cations are equivalent with respect to the substrate
atom environment. In contrast, only two sites of the four Cl anions
are equivalent and are differentiated by assigning different labels
to each inequivalent Cl anion, as shown in Fig. \ref{fig:PC_cell_fit}.
Furthermore, we find that the geometrical relaxations of the Cu substrate
atoms are small compared to the bare surface. In fact, the standard
deviation of the $z$ coordinates for the displacements of the Cu
atoms of the outer metal layer is about 0.04 \AA{}. Thus, we can assume
that all Cu atoms of the outer layer are located in the same plane,
and use the image plane position $z_{\mathrm{im}}$ of the bare substrate
to approximate the electrostatic response of metal substrate. The
details of the calculations of $z_{\mathrm{im}}$ for the bare Cu(100)
surface using a slab representation of the surface is described in
\ref{app:image_plane}. We find that $z_{\mathrm{im}}$ is converged
when increasing the number of Cu layers to 13 and $z_{\mathrm{im}}$=
1.48 \AA{} with respect to the plane of the outermost Cu layer $(z\text{=0)}$.

\begin{figure}
\centering \includegraphics[scale=0.3]{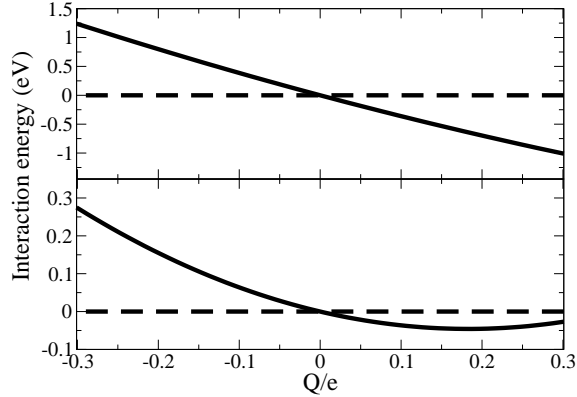}
\caption{Quadratic fits to calculated interaction energies $E^{int}[Q]=E[Q]-E[Q=0]$
(Upper panel) for a single surface unit cell of NaCl(2ML)/Cu(100)
and $\bar{E}_{\mathrm{PC}}^{int}[Q]=\bar{E}_{PC}[Q]-\bar{E}_{PC}[Q=0]$
(Lower panel) for a single surface unit cell of NaCl(2ML)/PC both
interacting with an external, uniformly charged plane as a function
of its charge $Q$. }

\label{fig:De_repfull} 
\end{figure}

The next step is to determine the effective work function $\Phi_{PC}$
in the PC model and the reference energy $\Delta E_{0}(\sigma)$.
To this end, we will use the reference geometry for the IF/M to be
the equilibrium geometry above for the primitive surface unit cell.
For this reference geometry, we will compute the interaction energy
$E^{\mathrm{int}}[Q]=E[Q]-E[Q=0]$ of the NaCl bilayer on the Cu(100)
slab with an external, uniformly charged plane with charge $Q=-e\sigma$,
and the corresponding interaction energy $\bar{E}_{\mathrm{PC}}^{\mathrm{int}}[Q]=\bar{E}_{PC}[Q]-\bar{E}_{PC}[Q=0]$
in the PC model of the Cu(100) slab. As shown in Fig. \ref{fig:De_repfull},
the computed interaction energies as a function of $Q$ are very well
approximated by a quadratic function, 
\begin{equation}
F[Q]=-W\frac{Q}{e}+\frac{Q^{2}}{2C}\ .\label{eq:quadfit}
\end{equation}
Here the linear term $W\delta Q/e$ corresponds to the energy required
to transfer an infinitesimal charge $\delta Q$ from the system to
the vacuum region and $C$ is the capacitance. In the full DFT calculation,
$W_{F}$ is simply to equal to the work function $\Phi$, since the
external plane is charged from the Fermi level of the metal substrate.
From the quadratic fit to the computed $E^{int}[Q]$ we obtain $W_{F}=3.74$
eV. This value is indeed very close to the calculated value for $\Phi$=
3.73 eV obtained from the computed Fermi energy with respect to the
vacuum level. \\
 In contrast, $W_{PC}$ differs from $\Phi$ in the PC approximation,
since the derivation of this approximation is based on charging from
the vacuum level. In this latter case, as shown in \ref{app:FOC_PC},
$W_{PC}$ is given by the electrostatic potential energy difference
$e\Delta\phi$ between the vacuum level and the PC plane. Here, the
extracted value for $W_{PC}=0.501$ eV from the corresponding fit
to $\bar{E}_{\mathrm{PC}}^{int}[Q]$ is close to the calculated value
$e\Delta\phi=0.526$ eV. The effective work function in the PC model
is then given by $\Phi_{PC}=W_{F}-W_{PC}=3.74-0.50=3.24$ eV.

The calculated capacitance $C_{PC}=\mathrm{\mathrm{0.365}}$ e/V in
the PC model is about 8\% smaller than the capacitance $C_{F}=\mathrm{\mathrm{0.395}}$
e/V in the full DFT calculation and could be corrected by adjusting
the position of the perfect conductor plane but that has not been
attempted here. Now, using Eq. (\ref{eq:DelEDef}) and the quadratic
form for the interaction energies in Eq. (\ref{eq:quadfit}), the
reference energy per primitive surface unit cell $\Delta E_{0}(\sigma)$
is given by, 
\begin{equation}
\Delta E_{0}(\sigma)=\Delta E_{0}+\left[\frac{1}{C_{F}}-\frac{1}{C_{PC}}\right]\frac{(e\sigma)^{2}}{2}.\label{eq:Eref0_Res}
\end{equation}
The linear term in $\sigma$ vanishes since $\Phi_{PC}=W_{F}-W_{PC}$.

\begin{figure}[t]
\centering %
\begin{tabular}{|c|c|}
\hline 
\includegraphics[scale=0.7]{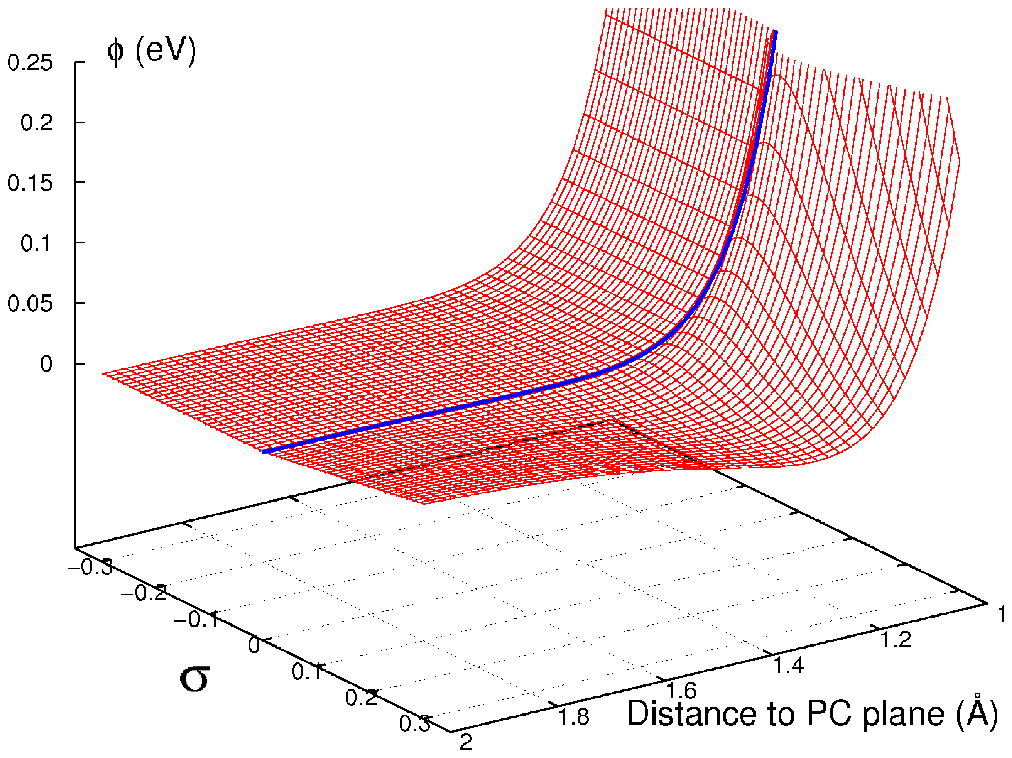}  & \includegraphics[scale=0.7]{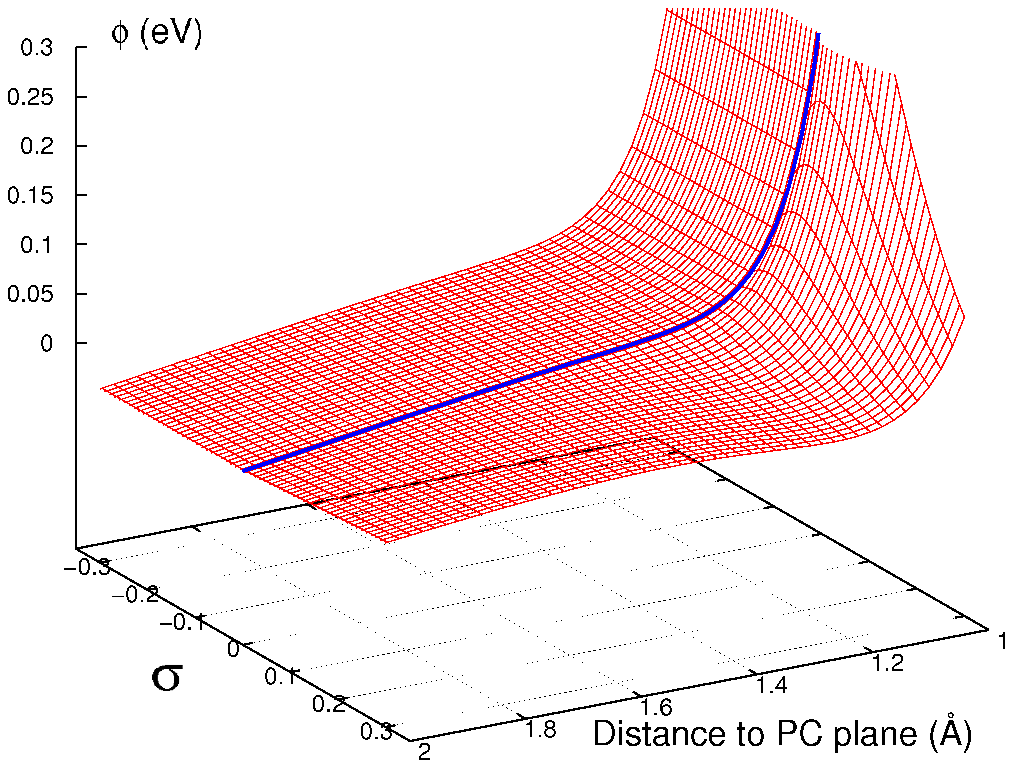}\tabularnewline
\hline 
Cl anion at site \textbf{A}  & Cl anion at site \textbf{B}\tabularnewline
\hline 
\includegraphics[scale=0.7]{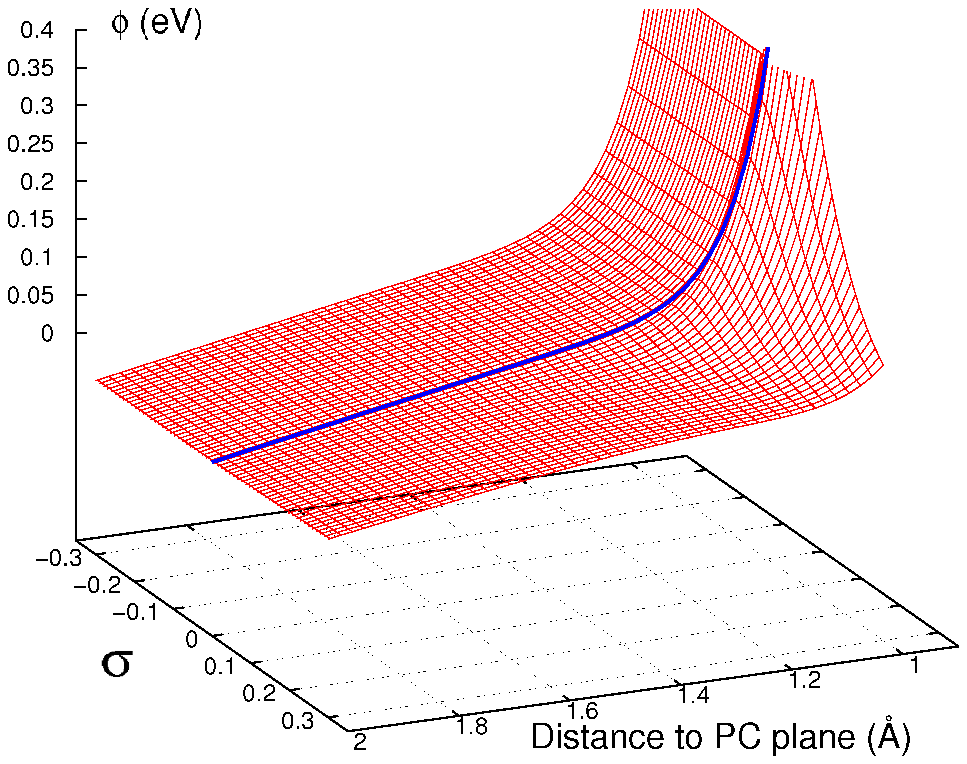}  & \includegraphics[scale=0.7]{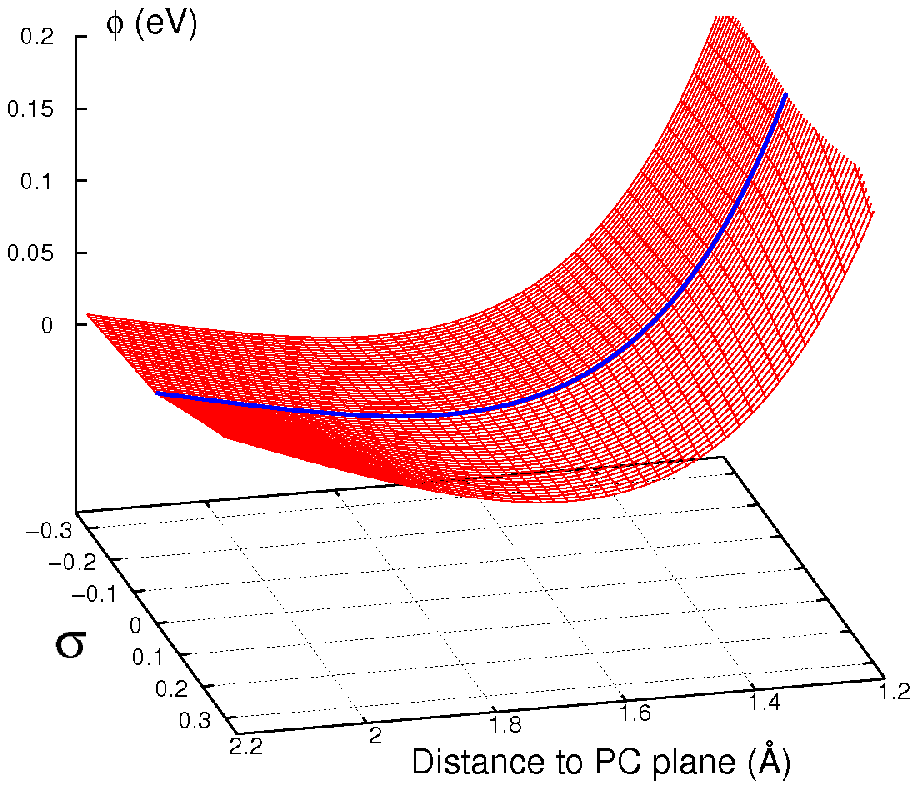}\tabularnewline
\hline 
Cl anion at site \textbf{C}  & Na cation\tabularnewline
\hline 
\end{tabular}\caption{(colour online) Calculated potentials $\phi$ as a function of the
atom distance $z$ to the PC plane and the charge$-e\sigma$ per surface
unit cell, for each atom of the NaCl layer as labelled in Fig. \ref{fig:PC_cell_fit}
(Right). The solid lines are the results for $\sigma=0$.}

{\small \label{fig:reppot_NaCl} } 
\end{figure}

The final step is to determine the potentials $\phi_{k}(z_{\mathrm{k}},\sigma)$
in the force field in Eqn.(\ref{eq:EPC_FF}) from how the energy $\Delta E(\sigma)$
in Eqn.(\ref{eq:DelEDef}) changes for the atoms of the NaCl bilayer
with respect to the their distances to the PC plane and $\sigma$.
Here, we have assumed that these potentials only affects the atoms
of the nearest layer to the metal surface, and has a dependence on
the atom kind and its atomic site. The inequivalent sites of the Cl
anions and Na cations with respect to substrate atoms, were identified
according to the labelling of Fig. \ref{fig:PC_cell_fit}. The calculation
of $\Delta E(\sigma)-\Delta E_{0}(\sigma)$, for different values
of $z_{\mathrm{k}}$ and $\sigma$ leads to various energy profiles
that decay rapidly with the distance to the PC plane. To fit the computed
set of data, we have used Morse functions 
\begin{equation}
\phi_{\mathrm{k}}(z_{\mathrm{k}},\sigma)=A_{\mathrm{k0}}\left[1-e^{-A_{\mathrm{k1}}(z_{\mathrm{k}}-A_{\mathrm{k2}})}\right]{}^{2}+A_{\mathrm{k3}}\label{eq:morse}
\end{equation}
where each coefficient $A_{\mathrm{k,i}}$ is at most a quadratic
function of $\sigma$. Figure \ref{fig:reppot_NaCl} shows the results
of the fitting of $\phi$ for each atom, as a function of $\sigma$
and the distance $z_{\mathrm{k}}$ to the PC plane inside the surface
unit cell. In addition, we show by the solid line the potential $\phi$
for $\sigma$=0. As expected, we find that the potential increases
when either the Na or Cl atom approaches to the image plane, and has
a weak dependence on the charge $e\sigma$ induced at the PC. In fact,
the presence of this charge will polarise the NaCl bilayer, such that
the ions will relax to a slightly different configuration. The use
of Morse-like functions to fit the data indicates that the potentials
are not purely repulsive, but they also exhibit small attractive contributions,
mainly for $z_{\mathrm{k}}$ close to their equilibrium values. In
the following, we will refer to $\phi_{\mathrm{k}}(z_{\mathrm{k}},\sigma)$
and $\phi_{\mathrm{k}}(z_{\mathrm{k}},\sigma=0)$ as polarised and
non-polarised potentials, respectively. 

\subsection{Charge states of an Au adatom on a NaCl bilayer supported by a Cu(100)
surface}

\label{subsec:chg_states} Our proposed approximate DFT-PC-FF method
will now be illustrated and tested by presenting results for the neutral
and negatively charged Au adatom on a NaCl bilayer supported by a
Cu(100) substrate. The quality of this method is judged by comparing
calculated values of adsorption energies and relaxed structures with
available results from DFT calculations that include the explicit
Cu(100) substrate, from now on referred to as DFT-FULL. Furthermore,
we present results for the vertical transition energy for charging
the Au adatom and the associated reorganisation energy, which cannot
be obtained from DFT-FULL calculations. For the DFT-PC-FF simulations,
we have considered supercells composed of $2\times2$ and $3\times3$
primitive surface units cells, being this surface unit cell as previously
defined in Fig. \ref{fig:PC_cell_fit}. For the DFT-FULL computations
we have only considered supercell with the $2\times2$ primitive surface
unit cells.

\begin{figure}
\centering \includegraphics[scale=0.09]{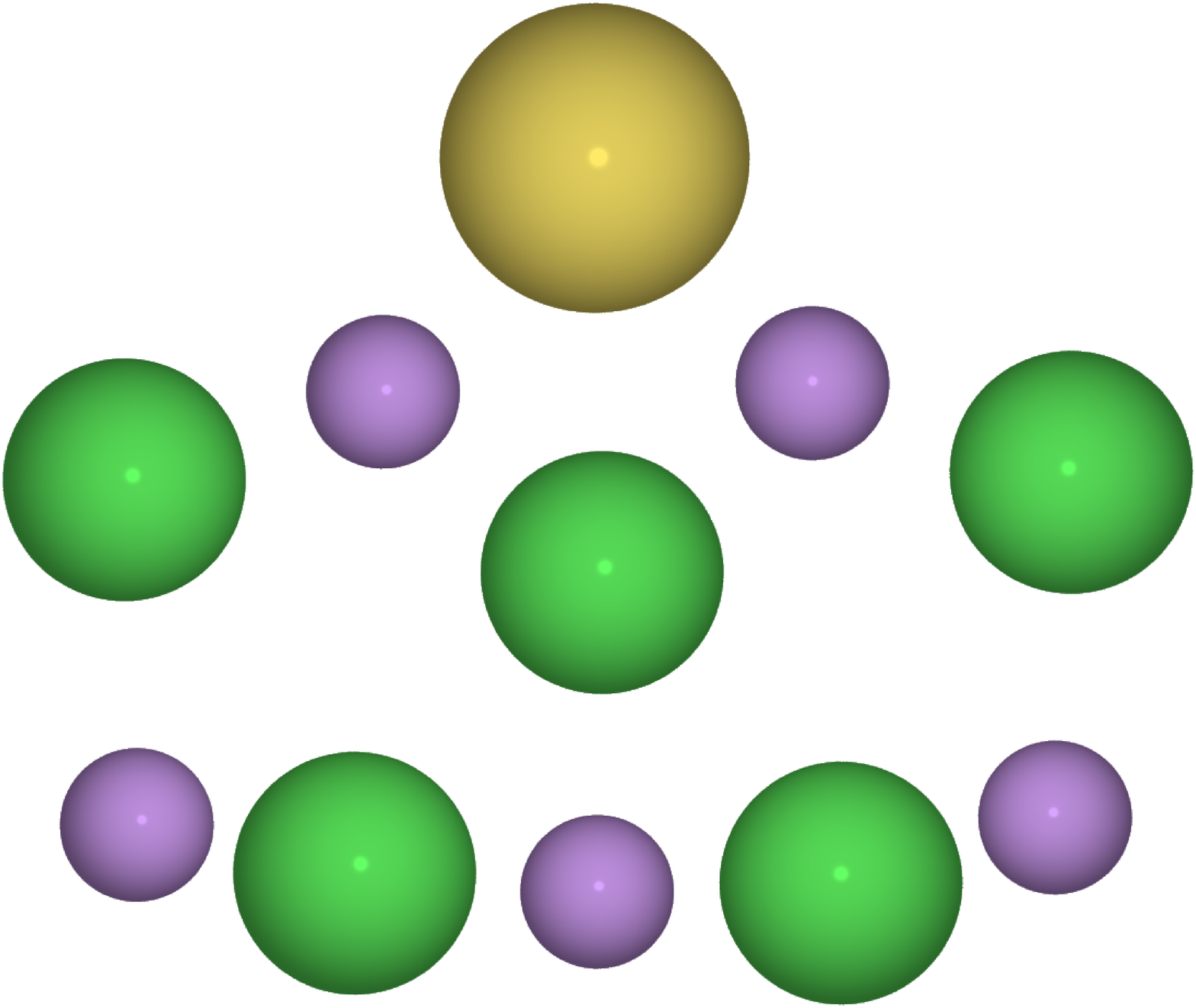}\hspace{1.2cm}
\includegraphics[scale=0.09]{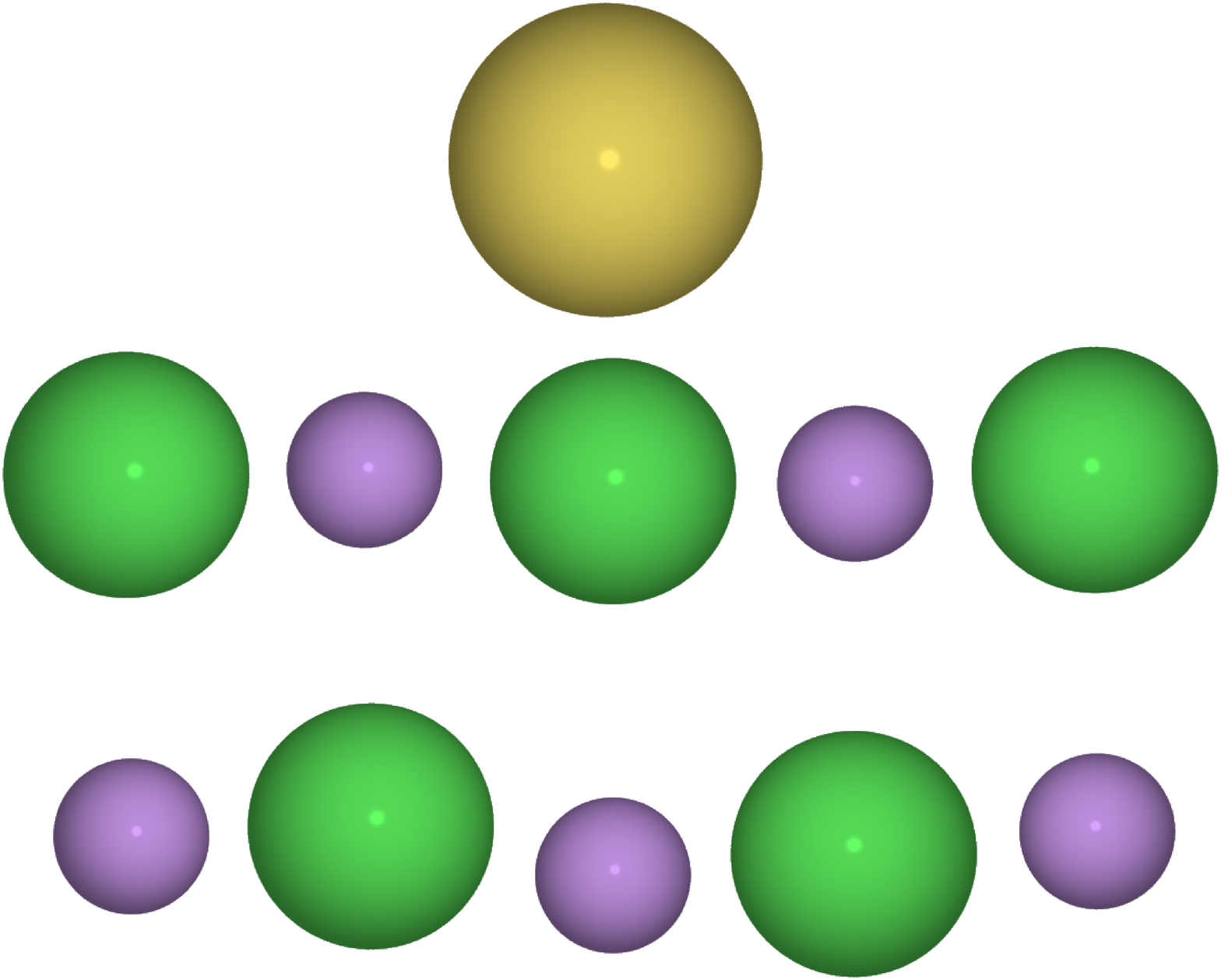} \caption{(colour online) Side view of the computed geometries for the negative
(Left) and the neutral (Right) Au atom (yellow) and Cl (green) and
Na (violet) ions. For both charge states, the Au adatom is situated
in top of a Cl anion. Whereas the negative Au (Au$^{-}$) adatom results
in large relaxations of the NaCl bilayer, the neutral charge state
(Au$^{0}$) of the Au adatom leaves the bilayer almost unaffected.}

\label{fig:Au_relaxed} 
\end{figure}

Previous STM experiments combined with DFT calculations showed that
the Au adatom has two possible charge states, either being negatively
charged or neutral, and each charge state result in a very different
ionic relaxations of the NaCl film \cite{repp3}. The ionic relaxation
in the DFT-FULL calculation, where only the two bottom Cu layers are
fixed, results in a negatively charged Au adatom situated on top of
a Cl anion and a large ionic relaxations of the NaCl layer, as schematically
shown in Fig. \ref{fig:Au_relaxed} (Left). Due to the electrostatic
interactions, the Cl anion coordinated to the Au anion is pushed towards
the bottom NaCl layer, while the four surrounding Na cations ions
are pulled outwards, in agreement with the results in Ref. \cite{repp3}.
In contrast to the large ionic relaxations of the NaCl bilayer, we
find that the relaxations of the Cu substrate atoms are small and
negligible. In fact, when keeping the Cu substrate atoms at their
positions in the absence of an adsorbate, we find that the resulting
total energy is only 0.03 eV larger than the total energy when the
two outer most Cu(100) layers are allowed to relax. The negligible
role of the small substrate relaxations justifies our simple force
field model with potentials that only depend on the distance from
the image plane position. Henceforth, all the DFT-FULL calculations
have been carried out by fixing the Cu(100) layers to their equilibrium
positions of the bare NaCl bilayer.

\begin{table}[ht]
\centering %
\begin{tabular}{|c|c|c|c|c|}
\hline 
 & DFT-FULL  & DFT\cite{repp3}  & \multicolumn{2}{c|}{DFT-PC-FF}\tabularnewline
\hline 
supercell size & $2\times2$  & $2\times2$  & $2\times2$  & $3\times3$\tabularnewline
\hline 
 & \multicolumn{4}{c}{\textbf{Au$^{-}$ }}\tabularnewline
\hline 
$d_{\mathrm{Au-Cl}}$(\AA{})  & 3.34  & 3.4  & 3.32 (3.35)  & 3.31 (3.32)\tabularnewline
\hline 
$E_{\mathrm{ads}}$(eV)  & 1.37  & 1.1  & 1.20 (1.27)  & 1.21 (1.28)\tabularnewline
\hline 
 & \multicolumn{4}{c}{\textbf{Au$^{0}$} }\tabularnewline
\hline 
$d_{\mathrm{Au-Cl}}$(\AA{})  & --  & 3.2  & 2.53  & 2.53\tabularnewline
\hline 
$E_{\mathrm{ads}}$(eV)  & --  & 0.4  & 0.64  & 0.68 \tabularnewline
\hline 
\end{tabular}\caption{Calculated adsorption energies $E_{\mathrm{ads}}$ and the distance
$d_{\mathrm{Au-Cl}}$ from the Au atom to the Cl anion for each charge
state of the Au adatom, when including the explicit Cu(100) substrate
in the DFT calculations (DFT-FULL), previous DFT results \cite{repp3}
and when using the new DFT-PC-FF method. Values in parenthesis show
the resulting values when using the polarisable potentials $\phi_{\mathrm{k}}(z_{\mathrm{k}},\sigma_{s})$
in the force field.}

\label{t:chg_Au} 
\end{table}

We now turn to the computation of the negatively charged Au adatom
using the DFT-PC-FF method. This state is now simply obtained by adding
one electron to the NaCl bilayer and the Au adatom. The added electron
will induce an equal but opposite charge at the PC plane, which ensures
the neutrality of the supercell. The resulting values for the adsorption
energy and the distance between the Au atom and the Cl atom underneath
are shown in Table \ref{t:chg_Au}. Results for DFT-PC-FF with the
polarisable force field are shown in parenthesis. In agreement with
the DFT-FULL calculations, we obtain a very similar relaxed geometric
structure. In fact, for the $2\times2$ supercell a DFT-FULL calculation
using the relaxed geometrical structure from the DFT-PC-FF calculation
with the non-polarisable force field gives an adsorption energy of
1.26 eV, which differ only by 0.11 eV from the adsorption energy 1.37
eV for the fully relaxed geometrical structure in the DFT-FULL calculation.
Instead, if we use the relaxed structure obtained with the polarisable
force field then the difference in the adsorption energy reduces from
0.11 eV to 0.04 eV. Furthermore, a comparison with the results from
the DFT-FULL calculations shows that DFT-PC-FF gives a minor difference
in the Au-Cl interatomic distance, always smaller than 0.02 \AA{},
independently if the force field is polarisable or not. Finally, the
adsorption energy is underestimated by about 0.17 eV (14\%) compared
to DFT-FULL when using the non-polarisable potentials $\phi_{\mathrm{k}}(z_{\mathrm{k}})$,
but is underestimated by only 0.10 eV (8\%) when using the polarisable
potentials $\phi_{\mathrm{k}}^{\mathrm{}}(z_{\mathrm{k}},\sigma)$.
Note that there is a huge reduction in the computational time with
about two orders of magnitude in the DFT-PC-FF calculations compared
to the DFT-FULL calculations.

The negatively charge state for the Au adatom is readily obtained
in our DFT-FULL calculations, whereas we have not been able to identify
a neutral state. In the DFT calculations reported in \cite{repp3},
they were able to identify a neutral-like state that had a small fractional
charge. In contrast, the DFT-PC-FF method provides a simplified and
efficient way to compute all those charge states that standard DFT
has a hard time or fails to predict. To obtain the neutral charge
state of the Au adatom in the DFT-PC-FF method, we simply set the
charge of the NaCl bilayer and the Au adatom to zero. In contrast
to the negative Au atom, we find that the NaCl bilayer is almost unaffected
by the presence of the neutral Au adatom, as schematically shown in
Fig. \ref{fig:Au_relaxed} (Right), in agreement with the earlier
calculations in Ref. \cite{repp3}. In comparison with the negative
charge state of the Au adatom, the neutral Au adatom is 0.8 \AA{}
closer to the Cl anion and has a smaller adsorption energy. Although,
the presented results are in agreement with those published in Ref.
\cite{repp3}, we obtain some differences in the calculated values,
because van der Waals interactions were included in our calculations.
Finally, note that the results in DFT-PC-FF are essentially converged
for the $2\times2$ supercell since they are very close to the results
for the $3\times3$ supercell.

\begin{figure}
\centering \includegraphics[scale=0.4]{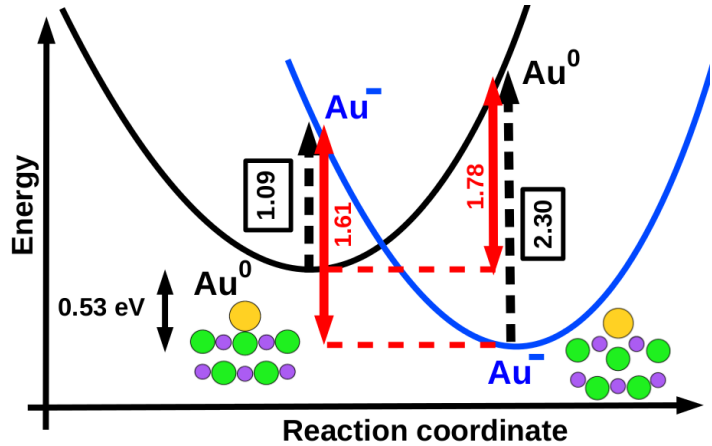} \caption{(colour online) Simple Marcus picture of charging and discharging
of the Au adatom on the NaCl bilayer supported by a Cu(100) surface.
The parabola indicate schematically the diabatic potential energy
curves as a function of the reaction coordinate. The vertical transition
energies (dashed lines) and reorganisation energies (solid lines)
are also indicated as calculated using the DFT-PC-FF method. The results
were obtained for a $3\times3$ supercell and a non-polarisable potential.
\label{fig:ScheMarcus}}
\end{figure}

In the STM experiments for the charging of an Au adatom on a NaCl
bilayer supported by a Cu(111) surface, an analysis of the observed
switching rate with bias suggested that the charging occurred by tunnelling
electron attachment to a Au$^{-}$ adatom state at about 1.4 eV. This
vertical transition energy is given by the energy to charge the Au
adatom in the equilibrium structure of the NaCl bilayer with the neutral
Au adatom. This anionic state cannot be realised in the DFT-FULL calculations
since it is an electronically excited state for this structure. However,
this anionic state is an electronic ground state in the DFT-PC-FF
scheme in the presence of an extra electron. Furthermore, with DFT-PC-FF
we can also calculate the vertical transition energy for neutralising
the Au$^{-}$ adatom given by the energy difference between the neutral
Au adatom and the Au$^{-}$ adatom in the equilibrium structure of
the NaCl bilayer and the Au adatom. In Fig. \ref{fig:ScheMarcus}
, we show the calculated values for these two transition energies
in a diagram that also shows the similarities of this charging and
discharging mechanisms to the classical Marcus picture for electron
transfer with schematic diabatic potential energy curves. From the
calculated transition and the adsorption energies of the neutral and
charged adatom in Table \ref{t:chg_Au}, we also obtain directly the
reorganisation energies in this picture associated with the geometrical
relaxations of NaCl bilayer for charging and discharging. The calculated
values of theses energies are also shown in Fig. \ref{fig:ScheMarcus}
and are rather close to each other. In fact, these energies should
be equal in a simple linear ionic and electron response model for
the NaCl bilayer on the Cu substrate to the adatom charge.

In order to compare the calculated transition energy for charging
of the Au adatom with the experimental value, we need to correct for
the workfunction difference for the supported NaCl bilayer when changing
the substrate from Cu(100) to Cu(111). This workfunction difference
is about 0.35 eV larger than when using the Cu(111) substrate \cite{penta},
so according to Eq.(\ref{eq:EPC_FF}), we have to shift the energies
of the Au$^{-}$ adatom in Fig. upwards with about 0.35 eV. This changes
the transition energy for charging of the Au adatom from 1.09 eV to
1.44 eV in close agreement with the value of 1.4 eV suggested by experiments.

\section{Concluding remarks}

\label{sec:conclusions}

The applicability of periodic density functional theory (DFT) methods
to calculate the total energy and forces of charged adsorbates on
ultra-thin insulating films, supported by a metal substrate, are severly
limited by the inherent delocalisation error of current exchange-correlation
functionals. Here, we have developed and presented a simplified and
efficient DFT method that surmounts this limitation. This new DFT-PC-FF
method is based on the perfect conductor (PC) model to approximate
the electrostatic response of the metal substrate, while the film
and the adsorbate are both treated fully within DFT. The missing interactions
between the metal substrate and the insulating film in the PC model
are modelled by a simple force field (FF). The parameters of the PC
model and the force field are obtained from DFT calculations of the
film and the substrate. In order to obtain some of these parameters
and the polarisability of the force field, we have to include an external,
uniformly charged plane in the DFT calculations, which has required
the development of an extension of periodic DFT to include such a
charged plane within a supercell. An extension that should be of more
general interest and applicable to other challenging problems, for
instance, in electrochemistry. The developed DFT-PC-FF method allows
us to handle the different charge states of adsorbates in a controlled
manner. Another most important advantage of this new scheme is the
large reduction in computer time and memory , since the metal electrons
are not explicitly included in the calculation.

The proposed DFT-PC-FF method is illustrated and tested by considering
the specific case of a NaCl bilayer, which is supported by a Cu(100)
substrate. We have carried out calculations for neutral and charged
Au adatoms on this film and compared the results with results from
DFT calculations that explicitly include the Cu(100) substrate, although
such a comparison was not possible for every charge state. In addition,
we have calculated the vertical transition energy for charging the
Au adatom and obtain a close agreement with the value suggested by
experiments. These energies cannot be obtained from DFT calculations
that include the full metal substrate.

Our results show that the DFT-PC-FF method not only predicts encouraging
results for adsorption and transition energies and relaxed structures
of charged adsorbates, but also reduces considerably the computational
time by a factor of almost two orders of magnitude. In fact, the possibility
to perform efficient DFT simulations by controlling the charge state
of adsorbates will allow to study various physical processes and properties,
which are currently either extremely challenging or not possible due
to the charge delocalisation error. In this respect, some interesting
problems involving insulating polar films we plan to address in the
near future are the following: (1) calculation of diffusion barriers
for adsorbates in various charge states; (2) HOMO-LUMO gaps of molecular
complexes; (3) molecular dynamics simulations of bond formation and
breaking upon charging and discharging.

\ack

The authors acknowledge Leverhulme Trust for funding this project
trough the grant (F/00 025/AQ) and allocation of computer resources
at HECToR through the membership in the materials chemistry consortium
funded by EPSRC (EP/F067496) and at Lindgren, PDC through SNIC. Mats
Persson is grateful for the support from the EU project ARTIST.

\appendix

\section{Calculation of the image plane for Cu(100)}

\label{app:image_plane} 
\begin{figure}
\centering \includegraphics[scale=0.11]{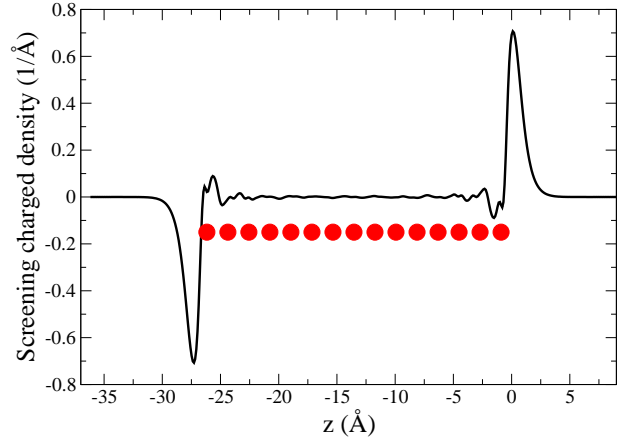}\caption{(colour online) Laterally average of induced charge density $\rho_{\mathrm{ind}}({\bf r})$
along the $z$ direction\label{fig:LatAvIndCh}. }
\end{figure}

\begin{figure}
\centering \includegraphics[scale=0.11]{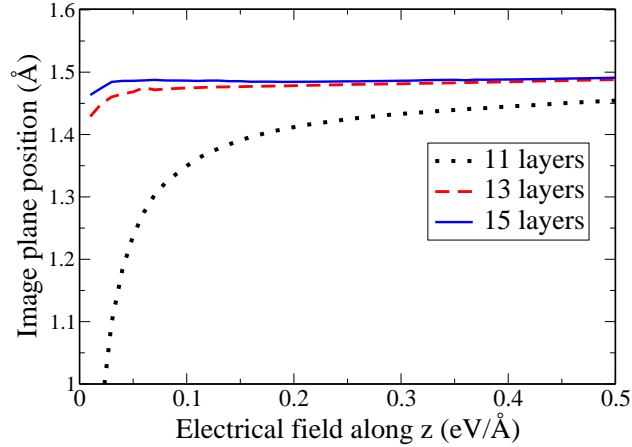} \caption{(colour online) Position of the image plane as a function of the applied
electrical field and for slabs with different number of Cu layers.\label{fig:ImPlanPos} }

\label{fig:image_vs_ef} 
\end{figure}

The calculations of the position $z_{im}$
of the image plane follows the classical work by Lang and Kohn \cite{kohn_lang3},
where this position with respect to outer most surfaces layer is given
by the centroid of the induced charged density $\rho_{\mathrm{ind}}({\bf r})$
by an external homogeneous electric field as, 
\begin{equation}
z_{\mathrm{im}}=\frac{\int z\ \rho_{\mathrm{ind}}({\bf r})d{\bf r}}{\int\rho_{\mathrm{ind}}({\bf r})d{\bf r}}+\frac{D}{2},\label{eq:zim}
\end{equation}
 where $D$ is the interlayer separation. In this work, $\rho_{\mathrm{ind}}({\bf r})$
has been calculated using VASP using the standard implementation of
an external, homogeneous field by an external surface dipole layer.
To model the Cu surface, we have used a slab geometry and gone up
to 15 Cu layers with a vacuum region of 20 \AA{}. Each layer contains
a single Cu atom and the interatomic distance was set to the calculated
bulk value of 2.546 \AA{}. The exchange-correlation effects were described
by the PBE functional and the ion-core interactions using the PAW.
The Brillouin zone was sampled by $11\times11\times1$ $k$ point
grid. Since the external electric field is applied on both sides of
the slab, $\rho_{\mathrm{ind}}({\bf r})$ becomes anti-symmetric,
as shown in Fig. \ref{fig:LatAvIndCh}, and the net induced charge
is zero. In order to approximate a semi-infinite surface, the centroid
of $\rho_{\mathrm{ind}}({\bf r})$ in Eq.(\ref{eq:zim}) was evaluated
by integrating $\rho_{\mathrm{ind}}({\bf r})$ to the centre of the
slab. As also shown in Fig. \ref{fig:LatAvIndCh}, the induced charge
density has weak, long-range Friedel oscillations into the bulk, so
it is necessary to increase the number of Cu layers to at least 13
to get a converged result for different strengths of the external
electric field (Fig. \ref{fig:ImPlanPos}). Note that the small deviations
in the region of small electrical fields for the slabs composed of
13 and 15 layers is expected and caused by numerical cancellation
errors in determining $\rho_{\mathrm{ind}}({\bf r})$.

\section{Modification of the energy functional in the presence of an external,
uniformly charged plane\label{app:ModEplan}}

\label{app:newDFT} In this section, we derive how the total energy
functional in density functional theory is modified for a system S
in the presence of an external, uniformly charged plane (UCP) with
a total charge $Q_{\mathrm{ext}}$ located at a position $z=z_{\mathrm{ext}}$.
The system S and the UCP is represented in a supercell and S is assumed
to include a metal substrate so that the total induced charge in S
is $Q_{s}=-Q_{\mathrm{ext}}$ and the super cell is neutral. Here,
$L_{\mathrm{z}}$ is the length of the supercell along the perpendicular
($z$) direction to the metal substrate and $A$ is its cross sectional
area. The electrostatic potentials of the charge density $\rho_{\mathrm{ext}}({\bf r})=\frac{Q_{ext}}{A}\delta(z-z_{\mathrm{ext}})$
of UCP and the charge density $\rho_{s}({\bf r})$ of the system S
in the supercell are denoted by $\phi_{\mathrm{ext}}({\bf r})$ and
$\phi_{\mathrm{s}}({\bf r})$, respectively. The UCP will only affect
directly the electrostatic part $E_{\mathrm{el,UCP}}[n_{\mathrm{s}}]$
of the total energy functional, whereas the kinetic energy and the
exchange correlation functionals remain unaffected. With this, $E_{\mathrm{el,UCP}}[n_{\mathrm{s}}]$
is given by 
\begin{eqnarray}
E_{\mathrm{el,UCP}}[n_{\mathrm{s}}]=\frac{1}{2}\int_{\mathcal{V}}\rho_{\mathrm{s}}({\bf r})\phi_{\mathrm{s}}({\bf r})d{\bf r}+\nonumber \\
\int_{\mathcal{V}}\rho_{\mathrm{s}}({\bf r})\phi_{\mathrm{ext}}({\bf r})d{\bf r}+\frac{1}{2}\int_{\mathcal{V}}[\rho_{\mathrm{s}}+\rho_{\mathrm{ext}}]({\bf r})\phi_{\mathrm{dip}}(z)d{\bf r}.\label{eq:Elec_full}
\end{eqnarray}

where the dependence on $Q_{\mathrm{ext}}$ has been indicated. Here
the first term on the RHS is the electrostatic self-interaction energy
of the charge density $\rho_{s}({\bf r})$, the second term is the
electrostatic interaction energy between $\rho_{s}({\bf r})$ and
$\rho_{\mathrm{ext}}({\bf r})$, and the last term is the dipole energy
correction. The latter dipole term corrects for the effects on the
electrostatic potential from the periodic boundary conditions in the
$z$ direction, as discussed by Neugebauer and Scheffler \cite{scheffler}
and later corrected by Bengtsson \cite{bengtsson}. The dipole potential
$\phi_{\mathrm{dip}}(z)$ is generated by a uniform surface dipole
layer with a surface dipole $-m$ determined by the total perpendicular
dipole moment of total charge distribution: 
\begin{equation}
m=\frac{1}{A}\int_{\mathcal{V}}\rho_{\mathrm{s}}({\bf r})zd{\bf r}+\frac{z_{\mathrm{ext}}Q_{ext}}{A}.\label{eq:dipA_full}
\end{equation}
 In the case when the dipole layer is located at $z=L_{z},$ the dipole
potential is given by, 
\begin{equation}
\phi_{\mathrm{dip}}(z)=4\pi m\left[\frac{z}{L_{\mathrm{z}}}-\frac{1}{2}\right]\ ,\ 0<z<L_{\mathrm{z}}\ .\label{eq:dippot}
\end{equation}
 The functional derivative of Eq. (\ref{eq:Elec_full}) with respect
to the electronic density $n_{\mathrm{s}}({\bf r})$ gives the dipole-corrected
electrostatic part $\phi_{\mathrm{UCP}}({\bf r})$ of the K-S potential
\begin{eqnarray}
\phi_{\mathrm{UCP}}({\bf r})=\phi_{\mathrm{s}}({\bf r})+\phi_{\mathrm{ext}}({\bf r})+\phi_{\mathrm{dip}}(z).\label{eq:phi_el_full_Def}
\end{eqnarray}
 Thus, the only modification of the K-S potential is that its electrostatic
part is replaced by $\phi_{\mathrm{UCP}}({\bf r})$ given in Eq.(\ref{eq:phi_el_full_Def}).
Furthermore, the same replacement needs to be done in the calculation
of the Hellman-Feynman forces. Usually, the kinetic energy is calculated
from the one-electron sum, which generates double counting terms.
In this case only the double counting term $E_{\mathrm{el,UCP}}^{\mathrm{DC}}$
from the electrostatic energy is modified and is given by, 
\begin{equation}
E_{\mathrm{el,UCP}}^{\mathrm{DC}}(Q_{\mathrm{ext}})=-\int_{\mathcal{V}}\rho_{\mathrm{e}}({\bf r})[\phi_{\mathrm{s}}({\bf r})+\phi_{\mathrm{ext}}({\bf r})+\phi_{\mathrm{dip}}(z)]\label{eq:DC_full_Def}
\end{equation}
 where $\rho_{\mathrm{e}}({\bf r})=-en_{\mathrm{s}}({\bf r})$ . Adding
this term to the dipole-corrected electrostatic energy of Eqn. (\ref{eq:Elec_full}),
one obtains,

\begin{eqnarray}
[E_{el,UCP}+E_{\mathrm{el,UCP}}^{\mathrm{DC}}](Q_{\mathrm{ext}})=-\frac{1}{2}\int_{\mathcal{V}}\rho_{\mathrm{e}}({\bf r})\phi_{\mathrm{e}}({\bf r})d{\bf r}+\nonumber \\
\frac{1}{2}\int_{\mathcal{V}}\rho_{\mathrm{i}}({\bf r})\phi_{\mathrm{i}}({\bf r})d{\bf r}+\nonumber \\
\frac{1}{2}\int_{\mathcal{V}}[\rho_{\mathrm{i}}-\rho_{\mathrm{e}}-\rho_{\mathrm{ext}}]({\bf r})\phi_{\mathrm{dip}}(z)d{\bf r}+\nonumber \\
\int_{\mathcal{V}}\rho_{\mathrm{ext}}({\bf r})[\phi_{\mathrm{i}}({\bf r})+\phi_{\mathrm{dip}}(z)]d{\bf r}\label{eq:Elec-EDC_full}
\end{eqnarray}
 where $\phi_{\mathrm{i}}$ is the electrostatic potential from the
ionic charge charge density $\rho_{\mathrm{i}}({\bf r})$ in the supercell.
Note that the integrals over the supercell are carried out in reciprocal
space by excluding the ${\bf g}=0$ component.

\section{Modification of the PC energy functional in the presence of an external,
uniformly charged plane}

\label{app:ModEPCplan}

\label{subapp:newDFT_PC} In this Section, we derive how the total
energy functional is modified in the case of perfect conductor (PC)
model of the metal substrate and an external closed system S in the
presence of an external, uniformly charged plane (UCP) with a total
charge $Q_{\mathrm{ext}}$ located at a position $z=z_{ext}$. In
this case, the system S can have a net charge $Q_{s}$. For further
details of the formalism behind the PC model, we refer the reader
to Ref. \cite{ISMP}. As in the previous case only the electrostatic
contribution to the total energy functional and consequently the K-S
potential will be modified by the UCP. The presence of the UCP and
S will induce a charge density at the perfect conductor plane located
at $z=z_{\mathrm{PC}}$ and is defined as, 
\begin{equation}
\rho_{\mathrm{ind}}({\bf r})=\sigma_{\mathrm{ind}}({\bf R})\delta(z-z_{\mathrm{PC}}).\label{eq:rhoind}
\end{equation}
 The laterally-averaged, induced surface charge density $\bar{\sigma}_{\mathrm{ind}}$
screens completely the total charge of UCP and $S$, and is given
by, 
\begin{equation}
\bar{\sigma}_{\mathrm{ind}}=-\frac{(Q_{\mathrm{s}}+Q_{\mathrm{ext}})}{A},\label{eq:sigG0}
\end{equation}
 and the laterally varying part of the surface charge density $\sigma_{\mathrm{ind}}^{\prime}({\bf r})$
is determined by the electrostatic potential $\phi_{\mathrm{s}}({\bf r})+\phi_{\mathrm{ext}}({\bf r})$
from $\rho_{\mathrm{s}}({\bf r})+\rho_{\mathrm{ext}}({\bf r})$ and
in reciprocal space it is given by, 
\begin{equation}
\sigma_{\mathrm{ind}}({\bf G})=-\frac{G}{2\pi}[\phi_{\mathrm{s}}+\phi_{\mathrm{ext}}](z_{\mathrm{PC}},{\bf G})\ ,\label{eq:sigG}
\end{equation}
 for non-zero reciprocal lattice vectors (${\bf G}\neq0$) of the
supercell, following the notation of Ref. $\mathrm{\cite{ISMP}}$.
Note that the ${\bf G}=0$ component of $\sigma_{\mathrm{ind}}$ is
equal to $\bar{\sigma}_{\mathrm{ind}}$. The electrostatic energy,
$E_{\mathrm{el,UCP}}^{\mathrm{PC}}$, of the system $S$ interacting
with the PC and the external charge plane UCP is then given by, 
\begin{eqnarray}
 &  & E_{\mathrm{el,UCP}}^{\mathrm{PC}}(Q_{\mathrm{ext}})=\frac{1}{2}\int_{\mathcal{V}}\rho_{\mathrm{s}}({\bf r})\phi_{\mathrm{s}}({\bf r})d{\bf r}+\nonumber \\
 &  & \int_{\mathcal{V}}\rho_{\mathrm{s}}({\bf r})\phi_{\mathrm{u}}({\bf r})d{\bf r}+\frac{1}{2}\int_{\mathcal{V}}[\rho_{\mathrm{s}}+\rho_{\mathrm{ext}}]({\bf r})\phi_{\mathrm{dip}}(z)d{\bf r}+\nonumber \\
 &  & \frac{1}{2}\int_{\mathcal{V}}\rho_{\mathrm{ind}}({\bf r})[\phi_{\mathrm{s}}+\phi_{\mathrm{ext}}]({\bf r})d{\bf r}\label{eq:Elec_PC}
\end{eqnarray}
 whose form differs from Eqn. (\ref{eq:Elec_full}) by the electrostatic
interaction of $\rho_{\mathrm{ind}}({\bf r})$, with the the charge
of $S$ and UCP. The dipole potential $\phi_{\mathrm{dip}}(z)$ has
been previously defined in Eqn. (\ref{eq:dippot}) but the surface
dipole moment $m$ in Eq. \ref{eq:dipA_full} now contains also a
contribution from $\rho_{\mathrm{ind}}({\bf r})$, 
\begin{equation}
m=\frac{1}{A}\int_{\mathcal{V}}[\rho_{\mathrm{s}}+\rho_{\mathrm{ind}}]({\bf r})zd{\bf r}+\frac{z_{\mathrm{ext}}Q_{\mathrm{ext}}}{A}.\label{eq:dipA_PC}
\end{equation}
 Rearranging terms in Eqn. (\ref{eq:Elec_PC}), we obtain 
\begin{eqnarray}
E_{\mathrm{el,UCP}}^{\mathrm{PC}}(Q_{\mathrm{ext}})=\frac{1}{2}\int_{\mathcal{V}}\rho_{\mathrm{s}}({\bf r})[\phi_{\mathrm{s}}+\phi_{\mathrm{ind}}+\phi_{\mathrm{ext}}+\phi_{\mathrm{dip}}]({\bf r})d{\bf r} & +\nonumber \\
\frac{1}{2}\int_{\mathcal{V}}\rho_{\mathrm{u}}({\bf r})[\phi_{\mathrm{s}}+\phi_{\mathrm{ind}}+\phi_{\mathrm{ext}}+\phi_{\mathrm{dip}}]({\bf r})d{\bf r} & -\nonumber \\
\frac{1}{2}\int_{\mathcal{V}}\rho_{\mathrm{ext}}({\bf r})\phi_{\mathrm{ext}}({\bf r})d{\bf r}\label{eq:Elec_PC2}
\end{eqnarray}
 where $\phi_{\mathrm{ind}}$ is the electrostatic potential from
$\rho_{\mathrm{ind}}({\bf r})$ in the supercell. Following the derivation
in Ref. \cite{ISMP}, the dipole-corrected electrostatic potential
$\phi_{\mathrm{dip-corr}}({\bf r})=[\phi_{s}+\phi_{\mathrm{ind}}+\phi_{\mathrm{u}}+\phi_{\mathrm{dip}}]({\bf r})$
within the supercell is equal, up to a constant, to the electrostatic
potential from $\rho_{\mathrm{s}}({\bf r})$, $\rho_{\mathrm{ind}}({\bf r})$
and $\rho_{\mathrm{u}}({\bf r})$ in the absence of the periodic boundary
conditions in the direction perpendicular to the PC plane . Therefore,
the electrostatic potential for the system $S$, $\phi_{\mathrm{el,UCP}}^{PC}({\bf r})$,
is simply obtained by adding a constant $\phi_{1}$ to the dipole
corrected potential, 
\begin{equation}
\phi_{\mathrm{el,UCP}}^{\mathrm{PC}}({\bf r})=\phi_{\mathrm{dip-corr}}({\bf r})+\phi_{1}\label{eq:phi_el_PC_Def}
\end{equation}
 and since the lateral average $\bar{\phi}_{\mathrm{el,UCP}}^{\mathrm{PC}}({\bf r}z)$
of $\phi_{\mathrm{el,UCP}}^{\mathrm{PC}}({\bf r})$ has to be zero
inside the PC , $\phi_{1}$ is determined by the following condition
at the PC plane 
\begin{equation}
\phi_{1}=-[\bar{\phi}_{\mathrm{s}}+\bar{\phi}_{\mathrm{ind}}+\bar{\phi}_{\mathrm{ext}}+\bar{\phi}_{\mathrm{dip}}](z_{\mathrm{PC}})\label{eq:PBC_PC_phi}
\end{equation}
 Finally, replacing $\phi_{\mathrm{dip-corr}}({\bf r})$ by $\phi_{\mathrm{el,UCP}}^{\mathrm{PC}}({\bf r})$
in Eqn. \ref{eq:Elec_PC2}, one gets the corrected expression for
the electrostatic energy, 
\begin{eqnarray}
 &  & E_{\mathrm{el,UCP}}^{\mathrm{PC}}(Q_{\mathrm{ext}})=\frac{1}{2}\int_{\mathcal{V}}\rho_{\mathrm{s}}({\bf r})\phi_{\mathrm{s}}({\bf r})d{\bf r}+\nonumber \\
 &  & \int_{\mathcal{V}}\rho_{\mathrm{s}}({\bf r})\phi_{ext}({\bf r})d{\bf r}+\frac{1}{2}\int_{\mathcal{V}}[\rho_{\mathrm{s}}+\rho_{\mathrm{ext}}]({\bf r})\phi_{\mathrm{dip}}(z)d{\bf r}+\nonumber \\
 &  & \frac{1}{2}\int_{\mathcal{V}}\rho_{\mathrm{ind}}({\bf r})[\phi_{\mathrm{s}}+\phi_{\mathrm{ext}}]({\bf r})d{\bf r}+\frac{1}{2}\phi_{1}(Q_{\mathrm{s}}+Q_{\mathrm{ext}})\label{eq:Elec_PC_final}
\end{eqnarray}
 and its form differs from Eqn. (\ref{eq:Elec_PC}) by the term $\frac{1}{2}\phi_{1}(Q_{\mathrm{s}}+Q_{\mathrm{ext}})$.
Before closing this section, we present the expressions for the double
counting terms used to evaluate the total energy when the kinetic
energy is obtained from the one-electron sum. Note that adding a constant
to the K-S potential does not change the kinetic energy and, therefore,
$\phi_{1}$ does not need to be included in the K-S potential so the
electrostatic part of the double counting term is given by, 
\begin{equation}
E_{\mathrm{el,UCP}}^{\mathrm{DC,PC}}(Q_{\mathrm{ext}})=-\int_{\mathcal{V}}\rho_{\mathrm{e}}({\bf r})[\phi_{\mathrm{s}}+\phi_{\mathrm{ind}}+\phi_{\mathrm{ext}}+\phi_{\mathrm{dip}}]({\bf r})d{\bf r}\label{eq:EDCDef_PC}
\end{equation}
 Adding this term to the dipole-corrected, electrostatic potential
energy in Eqn.(\ref{eq:Elec_PC_final}), one obtains 
\begin{eqnarray}
[E_{\mathrm{el,UCP}}^{\mathrm{PC}}+E_{\mathrm{el,UCP}}^{\mathrm{\tiny{DC,PC}}}](Q_{\mathrm{ext}})=-\frac{1}{2}\int_{\mathcal{V}}\rho_{\mathrm{e}}({\bf r})\phi_{\mathrm{e}}({\bf r})d{\bf r}+\nonumber \\
\frac{1}{2}\int_{\mathcal{V}}\rho_{\mathrm{i}}({\bf r})\phi_{\mathrm{i}}({\bf r})d{\bf r}+\nonumber \\
\frac{1}{2}\int_{\mathcal{V}}[\rho_{\mathrm{i}}-\rho_{\mathrm{e}}({\bf r})-\rho_{\mathrm{ext}}-\rho_{\mathrm{ind}}]({\bf r})\phi_{\mathrm{dip}}(z)d{\bf r}+\nonumber \\
\frac{1}{2}\phi_{1}(Q_{\mathrm{s}}+Q_{\mathrm{ext}})+\int_{\mathcal{V}}\rho_{\mathrm{ext}}({\bf r})[\phi_{\mathrm{i}}({\bf r})+\phi_{\mathrm{dip}}(z)]d{\bf r}+\nonumber \\
\frac{1}{2}\int_{\mathcal{V}}\rho_{\mathrm{ind}}({\bf r})[\phi_{\mathrm{ext}}-\phi_{\mathrm{e}}+\phi_{\mathrm{i}}+\phi_{\mathrm{dip}}]({\bf r})d{\bf r}\label{eq:Elec-EDC_PC_const}
\end{eqnarray}
 Note that in absence of a PC plane $\rho_{\mathrm{ind}}({\bf r})=0$,
$Q_{\mathrm{s}}=-Q_{\mathrm{ext}}$ and Eqn. (\ref{eq:Elec-EDC_PC_const})
reduces to (\ref{eq:Elec-EDC_full}).

\section{DFT implementation to include an external, uniformly charged plane}

\label{app:compimp} The required computational modifications for
the PC model have already been implemented in the VASP code \cite{vasp}
and is described in Ref. \cite{ISMP}. In our new DFT scheme, the
inclusion of the repulsive energy and forces over the atoms in the
bottom layer of the insulating film is straightforward. The implementation
of the DFT method to handle an external, uniformly charged plane UCP,
$\rho_{\mathrm{ext}}({\bf r})=\sigma_{\mathrm{ext}}\delta(z-z_{ext})$,
is also rather straightforward. The electrostatic potential $\phi_{\mathrm{ext}}({\bf r})$
from $\rho_{\mathrm{ext}}({\bf r})$ is generated in standard manner
by solving Poisson's equation in reciprocal space. By adding $\rho_{\mathrm{ext}}({\bf r})$
to the electronic charge density $n_{s}({\bf r})$ allowed us to compute
the surface dipole moment that determines the dipole potential through
Eqn. (\ref{eq:dippot}), as well as the dipole energy correction (third
term of the RHS of Eqn. (\ref{eq:Elec-EDC_full})) using the standard
VASP dipole correction subroutine. However, in the case of the PC
model with an UCP, the surface dipole moment and the dipole energy
correction are computed by adding the $\rho_{ext}({\bf r})$ and $\rho_{\mathrm{ind}}({\bf r})$
to $n_{s}({\bf r})$. Finally, the fifth and sixth terms of the RHS
of Eqn. (\ref{eq:Elec-EDC_PC_const}) have been computed in reciprocal
space.

\section{First order correction for a system interacting with a Perfect Conductor
and an external charged plane}

\label{app:FOC_PC}

Here, we show that the first order correction to the electrostatic
energy $E_{\mathrm{el,UCP}}^{\mathrm{PC}}$ in the PC model to the
charge $Q_{\mathrm{ext}}$ of an external, uniformly charged plane
is given by the difference of the averaged potential of the neutral
system between the positions $z_{\mathrm{ext}}$ and $z_{\mathrm{PC}}$
of the external plane and the PC plane, respectively. Differentiating
the electrostatic energy expression of Eqn. (\ref{eq:Elec_PC_final})
with respect to $Q_{\mathrm{ext}}$, one obtains, 
\begin{eqnarray}
 &  & \frac{\partial E_{\mathrm{el,UCP}}^{\mathrm{PC}}}{\partial Q_{\mathrm{ext}}}(Q_{\mathrm{ext}})=\frac{1}{2A}\int_{\mathcal{V}}\delta(z-z_{\mathrm{ext}})[2\phi_{\mathrm{s}}({\bf r})d{\bf r}+\phi_{\mathrm{dip}}(z)]+\nonumber \\
 &  & \frac{1}{2A}\int_{\mathcal{V}}\frac{\partial\phi_{\mathrm{dip}}(z)}{\partial Q_{\mathrm{ext}}}(z)[\rho_{\mathrm{s}}+\rho_{\mathrm{ext}}]({\bf r})d{\bf r}+\frac{1}{2}\phi_{1}-\nonumber \\
 &  & \frac{1}{2A}\int_{\mathcal{V}}\delta(z-z_{\mathrm{PC}})\phi_{\mathrm{s}}({\bf r})d{\bf r}\label{eq:FOC_dEdQ}
\end{eqnarray}
 where the dipole potential $\phi_{\mathrm{dip}}(z)$ is defined in
Eqn. (\ref{eq:dippot}) and the constant $\phi_{1}$ is given by Eqn.
\ref{eq:PBC_PC_phi}. Expressing the surface dipole moment $m$ of
Eqn. (\ref{eq:dipA_PC}) as $m=m^{(0)}+m^{(1)}$ with $m^{(0)}=\frac{1}{A}\int_{\mathcal{V}}\rho_{\mathrm{s}}({\bf r})zd{\bf r}$
and $m^{(1)}=\frac{1}{A}\int_{\mathcal{V}}[\rho_{\mathrm{ind}}({\bf r})+\rho_{\mathrm{ext}}({\bf r})]zd{\bf r}$,
and the corresponding contributions $\phi^{(1)}(z)$and $\phi^{(1)}(z)$
to $\phi_{\mathrm{dip}}(z)$, one obtains $\frac{\partial m^{(1)}}{\partial Q_{ext}}=\frac{(z_{\mathrm{ext}}-z_{\mathrm{PC}})}{A}$
and 
\begin{equation}
\frac{\partial\phi_{\mathrm{dip}}}{\partial Q_{\mathrm{ext}}}(z)=\frac{4\pi(z_{\mathrm{ext}}-z_{\mathrm{PC}})}{A}\left[\frac{z}{L_{\mathrm{z}}}-\frac{1}{2}\right].\ 
\end{equation}
 Inserting $\frac{\partial\phi_{\mathrm{dip}}}{\partial Q_{\mathrm{ext}}}(z)$
in Eqn. (\ref{eq:FOC_dEdQ}), one gets after re-arranging terms, 
\begin{equation}
\int_{\mathcal{V}}\frac{\partial\phi_{\mathrm{dip}}}{\partial Q_{\mathrm{u}}}(z)[\rho_{\mathrm{s}}+\rho_{\mathrm{ext}}]({\bf r})d{\bf r}=[\phi_{\mathrm{dip}}^{(0)}(z_{\mathrm{ext}})-\phi_{\mathrm{dip}}^{(0)}(z_{\mathrm{PC}})+\phi_{\mathrm{dip}}^{(1)}(z_{\mathrm{ext}})].
\end{equation}
 and, 
\begin{eqnarray}
 &  & \frac{\partial E_{\mathrm{el,UCP}}^{\mathrm{PC}}}{\partial Q_{\mathrm{ext}}}=[\bar{\phi}_{\mathrm{s}}+\phi_{\mathrm{dip}}^{(0)}](z_{\mathrm{ext}})-[\bar{\phi}_{\mathrm{s}}+\phi_{\mathrm{dip}}^{\mathrm{(1)}}](z_{\mathrm{PC}})+\nonumber \\
 &  & \frac{1}{2}[\phi_{\mathrm{dip}}^{\mathrm{(1)}}(z_{\mathrm{ext}})-\bar{\phi}_{\mathrm{ind}}(z_{\mathrm{PC}})-\bar{\phi}_{\mathrm{ext}}(z_{\mathrm{PC}})].\label{eq:FOC_dEdQ_final}
\end{eqnarray}
 Since $m^{(1)}=0$ and $\phi_{\mathrm{dip}}^{(1)}=\phi_{\mathrm{ext}}=0$
when $Q_{\mathrm{ext}}=0$, and $\bar{\phi}_{\mathrm{ind}}=0$ for
a neutral system S, we finally obtain the desired result 
\begin{equation}
\left.\frac{\partial E_{\mathrm{el,UCP}}^{\mathrm{PC}}}{\partial Q_{\mathrm{ext}}}\right|_{Q_{\mathrm{ext}}=0}=[\bar{\phi}_{\mathrm{s}}+\phi_{\mathrm{dip}}](z_{ext})-[\bar{\phi}_{\mathrm{s}}+\phi_{\mathrm{dip}}](z_{\mathrm{PC}}).\label{eq:FOC_dEdQ_final0}
\end{equation}


\begin{thebibliography}{10}
\bibitem{endnote1}The super cell dimensions are $L_{x}$=$L{}_{y}$=
7.638 \AA{} and $L_{z}$= 31.20 \AA{}
\bibitem{ext_plane_1} If the external plane is sufficiently far from
the surface, the value of $\Phi$ will be equal to the work function
of the system.
\end{thebibliography}
\end{document}